\documentclass[fleqn,usenatbib]{mnras}

\usepackage{newtxtext,newtxmath}
\usepackage[T1]{fontenc}
\usepackage{ae,aecompl}

\usepackage{graphicx}	% Including figure files
\usepackage{amsmath}	% Advanced maths commands
\usepackage{xcolor}
\usepackage{ulem}
\newcommand{\GJout}[1]{}
\title[Anomalous gas in ESO 149-G003]{Anomalous gas in ESO 149-G003: A MeerKAT-16 View}
\author[G. I. G. J\'ozsa et al.]{\noindent Gyula I. G. J\'ozsa$^{1,2,3}$\thanks{E-mail:jozsa@ska.ac.za},
Kshitij Thorat$^{2,1,4}$,
Peter Kamphuis$^{5}$,
Lerato Sebokolodi$^{1,2,6}$,
\newauthor Eric K. Maina$^{2}$,
Jing Wang$^{7}$,
Dani\"elle L. A. Pieterse$^{8}$,
Paul Groot$^{8,9,10}$, \newauthor
Athanaseus J. T. Ramaila$^{1}$,
Paolo Serra$^{11}$,
Lexy A. L. Andati$^{2}$,
W. J. G. de Blok$^{12,13,14}$,\newauthor
Benjamin V. Hugo$^{1,2}$,
Dane Kleiner$^{11}$,
Filippo M. Maccagni$^{11}$,
Sphesihle Makhathini$^{2}$,\newauthor
D\'{a}niel Cs. Moln\'{a}r$^{11}$,
Mpati Ramatsoku$^{2,11}$,
Oleg M. Smirnov$^{2,1}$,
Steven Bloemen$^{8}$, \newauthor
Kerry Paterson$^{15}$, %ORCID is [0000-0001-8340-3486]
Paul Vreeswijk$^{8}$,
Vanessa McBride$^{16}$,
Marc Klein-Wolt$^{8}$, \newauthor
Patrick Woudt$^{9}$,  
Elmar K\"ording$^{8}$, 
Rudolf Le Poole$^{17}$,
Sharmila Goedhart$^{1,18}$, \newauthor
Sean S. Passmoor$^{1}$, 
Maciej Serylak$^{1,19}$,
Ralf-Jürgen Dettmar$^{5}$
\\
$^{1}$South African Radio Astronomy Observatory, 2 Fir Street, Black River Park, Observatory, Cape Town, 7925, South Africa\\
$^{2}$Department of Physics and Electronics, Rhodes University, PO Box 94, Makhanda, 6140, South Africa\\
$^{3}$Argelander-Institut f\"ur Astronomie, Auf dem H\"ugel 71, D-53121 Bonn, Germany\\
$^{4}$Department of Physics, University of Pretoria, Hatfield, Pretoria 0028, South Africa\\
$^{5}$Ruhr University Bochum, Faculty of Physics and Astronomy, Astronomical Institute, 44780 Bochum, Germany\\
$^{6}$National Radio Astronomy Observatory, 1003 Lopezville Rd, Socorro, NM 87801, USA\\
$^{7}$Kavli Institute for Astronomy and Astrophysics, Peking University, Beijing 100871, China\\
$^{8}$Department of Astrophysics/IMAPP, Radboud University, P.O. 9010, 6500 GL, Nijmegen, The Netherlands\\
$^{9}$Inter-University Institute for Data Intensive Astronomy, Department of Astronomy, University of Cape Town, Private Bag X3, Rondebosch 7701, South Africa\\
$^{10}$South African Astronomical Observatory, P.O. Box 9, 7935 Observatory, South Africa\\
$^{11}$INAF - Osservatorio Astronomico di Cagliari, Via della Scienza 5, I-09047 Selargius (CA), Italy\\
$^{12}$Netherlands Institute for Radio Astronomy (ASTRON), Postbus 2, 7990 AA Dwingeloo, the Netherlands\\
$^{13}$Dept.\ of Astronomy, Univ.\ of Cape Town, Private Bag X3, Rondebosch 7701, South Africa\\
$^{14}$Kapteyn Astronomical Institute, University of Groningen, Postbus 800, 9700 AV Groningen, The Netherlands\\
$^{15}$Center for Interdisciplinary Exploration and Research in Astrophysics (CIERA) and Department of Physics and Astronomy, Northwestern University,\\ 1800 Sherman Ave, Evanston, IL 60201, USA\\
$^{16}$IAU-Office For Astronomy for Development, P.O. Box 9, 7935 Observatory, South Africa\\
$^{17}$Leiden Observatory, Leiden University, P.O. Box 9513, NL-2300 RA Leiden, The Netherlands\\
$^{18}$Centre for Space Research, North-West University, Potchefstroom campus, Private Bag X6001, Potchefstroom, 2520, South Africa\\
$^{19}$Department of Physics and Astronomy, University of the Western Cape, Bellville, South Africa
}

\date{Accepted 2020 December 2. Received 2020 November 6; in original form 2019 September 18}

\pubyear{2019}

\begin{document}
\label{firstpage}
\pagerange{\pageref{firstpage}--\pageref{lastpage}}
\maketitle

\begin{abstract}
ESO 149-G003 is a close-by, isolated dwarf irregular galaxy. Previous observations with the ATCA indicated the presence of anomalous neutral hydrogen (\ion{H}{i}) deviating from the kinematics of a regularly rotating disc. We conducted follow-up observations with the MeerKAT radio telescope during the 16-dish Early Science programme as well as with the MeerLICHT optical telescope. Our more sensitive radio observations confirm the presence of anomalous gas in ESO~149-G003, and further confirm the formerly tentative detection of an extraplanar \ion{H}{i} component in the galaxy. Employing a simple tilted-ring model, in which the kinematics is determined with only four parameters but including morphological asymmetries, we reproduce the galaxy's morphology, which shows a high degree of asymmetry. By comparing our model with the observed \ion{H}{i}, we find that in our model we cannot account for a significant (but not dominant) fraction of the gas. From the differences between our model and the observed data cube we estimate that at least 7\%--8\% of the \ion{H}{i} in the galaxy exhibits anomalous kinematics, while we estimate a minimum mass fraction of less than 1\% for the morphologically confirmed extraplanar component. We investigate a number of global scaling relations and find that, besides being gas-dominated with a neutral gas-to-stellar mass ratio of 1.7, the galaxy does not show any obvious global peculiarities. Given its isolation, as confirmed by optical observations, we conclude that the galaxy is likely currently acquiring neutral gas. It is either re-accreting gas expelled from the galaxy or accreting pristine intergalactic material. \end{abstract}


%%%%%%%%%%%%%%%%%%%%%%%%%\newauthor%%%%%%%%%%%%%%%%%%%%%%%%%

%%%%%%%%%%%%%%%%% BODY OF PAPER %%%%%%%%%%%%%%%%%%

\section{Introduction}
How galaxies get their star forming material is a point of discussion in studies of galaxy evolution. Trivially, at some time galaxies acquire gas from the intergalactic medium (IGM) and transform it into stars. The question at which evolutionary stage they do so, how the gas enters the galaxies, and by which channels is still a matter of debate. Theory and observations point towards a mode of direct accretion from the IGM, which may still take place at redsift $z\,=\,0$ for star-forming galaxies over a large mass range, from giant spirals to dwarf galaxies \citep[][e.g.]{Sancisi2008,theodoro2014}. It has, however, proven to be exceedingly difficult to detect \citep{Sancisi2008,theodoro2014}.\\
\indent
The way in which Milky Way-sized galaxies can even maintain their current star formation rates (SFRs) is an open debate from both the theoretical \citep{Nelson2013,Nelson2015,Huang2019} and the observational perspective \citep{Sancisi2008}. With observed average SFRs of $\sim\, 3\,M_\odot\,\mathrm{yr}^{-1}$ \citep{Bothwell2011} in the local universe it is clear that a typical massive galaxy would use up its gas supply within a few Gigayears if the gas is not replenished or recycled \citep{Kennicut1994}. Besides the SFRs, continuous accretion of gas onto the disc of spiral galaxies can also serve to explain other observed phenomena, such as asymmetries of galactic discs \citep{vEymeren2011, Giese2016} and warping \citep{Garcia-Ruiz2002} or the chemical composition of the galactic discs \citep{Chiosi1980}. Up to now observations of ongoing accretion have remained elusive in the sense that not enough accretion events are observed to explain the required accretion rates \citep{Sancisi2008}. Hence, observing an ongoing cold gas accretion event can provide crucial insights into how cold gas accretes onto the gas discs of galaxies.\\
\indent
One way to identify a possible ongoing accretion event is through the observation of galaxies with peculiar gas kinematics \citep[e.g.][]{swaters_hi_1997,schaap_vertical_2000,lee_ngc_2001,fraternali_deep_2002,barbieri_extra-planar_2005,oosterloo_cold_2007,boomsma_hi_2008,heald_westerbork_2011,zschaechner_halogas:_2011,zschaechner_halogas:_2012,gentile_halogas:_2013,de_blok_halogas_2014,vargas_halogas_2017,pingel_gbt_2018,marasco_halogas_2019}. Such kinematics can be observed as deviations from the bulk velocity produced by the virialised rotation of the gaseous disc in the gravitational potential. This bulk rotation can be well described by tilted ring models (TRMs) \citep{Rogstad1974,Jozsa2007} of the gas distribution in the discs of spiral galaxies. \\
\indent
While accretion in large spiral galaxies is the subject of several studies \citep[e.g.][]{heald_westerbork_2011}, little is known about accretion in dwarf galaxies.
\indent

Other than their more massive siblings, the majority of dwarf galaxies are either not forming stars (dwarf Ellipticals, dEs), or forming stars at a low rate (dwarf Irregulars, dIs or dwarf LSB galaxies). dIs are forming stars very inefficiently compared to spiral galaxies \citep{brosch_star_1998,bothwell_interstellar_2009,roychowdhury_star_2009}. As a consequence their gas depletion time exceeds on average a Hubble time significantly ( \citealt{van_zee_evolutionary_2001} estimate a typical gas depletion time of $\sim$20~Gyr in a sample of $\sim$50 dIs). Some fraction of dwarf galaxies (about 6\% in the Local Volume), though, are currently in a starburst phase, and 25\% of all stars in dwarf galaxies (dI and dE) in the Local Volume are produced in starbursts \citep[][]{lee_dwarf_2009}. This coincides well with the fraction of 20\% of stars born in starbursts in the present-day general galaxy population (\citealt{brinchmann_physical_2004}, see also \citealt{bergvall_star_2012}). This implies that a star burst phase is a typical phenomenon for dwarf galaxies. Studies on gas accretion in dwarf galaxies are hence not about explaining how their SFRs are sustained but rather about how and when these starbursts are triggered.\\
\indent
Compared to dIs, starburst dwarf galaxies  (\ion{H}{ii} galaxies, \citealt{terlevich_spectrophotometric_1991,taylor_survey_1993}, or Blue Compact Dwarf (BCD) galaxies, \citealt{gil_de_paz_palomarcampanas_2003}) do not only show enhanced SFRs  \citep{bergvall_star_2012} but also have more concentrated
density profiles in their stellar \citep{papaderos_optical_1996-1,van_zee_neutral_1998,lelli_evolution_2014}  and \ion{H}{i} \citep{taylor_star_1994,van_zee_neutral_1998,lelli_evolution_2014, lelli_dynamics_2014} components. These concentrated density profiles are kinematically visible through the  steeper velocity gradients in their
rotation curves \citep{van_zee_complex_1998,van_zee_kinematic_2001,lelli_dynamics_2012a,lelli_dynamics_2012b}.
 Stronger radial color-gradients \citep{papaderos_optical_1996} are typically a result of a more centrally concentrated star formation \citep{taylor_star_1994,van_zee_evolutionary_2001}.
%The gaseous metallicity of BCDs is found to be lower than that of dIs and dEs \citep{terlevich_spectrophotometric_1991,izotov_new_1991,izotov_heavy-element_1999,izotov_heavy-element_1999,zhao_study_2013},. GJ: notice that while this is a statement from Verbeke et al. 2014, I could not find this to be confirmed in the literature.}
\\
\indent
It has hence been pointed out that the transition of dwarf galaxies from one type (dI, BCD, dE) to another (if it takes place) has to involve the redistribution of mass and angular momentum \citep{van_zee_kinematic_2001,bergvall_star_2012,lelli_dynamics_2012b,lelli_triggering_2014}. Potential mechanisms for this are tidal interaction \citep{taylor_star_1994,taylor_h_1995,taylor_h_1996,taylor_h_1997,van_zee_complex_1998,lelli_dynamics_2012a} and gas-rich mergers \citep{van_zee_complex_1998,lelli_dynamics_2012a,lelli_triggering_2014, papaderos_optical_1996-1,lelli_triggering_2014}, but also direct gas infall \citep{papaderos_optical_1996-1,lopez-sanchez_intriguing_2012,lelli_triggering_2014,ashley_hi_2017}. 
Simulations show that all of these mechanisms can play a role. Gas-rich mergers and tidal interactions of gas-rich galaxies \citep{bekki_formation_2008} as well as infall of gas clouds \citep{verbeke_gaseous_2014} are able to result in simulated systems reminiscent of BCDs.\\
\indent
While the literature is dominated by observational evidence for tidal interaction and mergers between gas-rich dIs as a potential formation scenario \citep{taylor_h_1995,taylor_h_1996,taylor_h_1997,van_zee_complex_1998,karachentsev_new_2007,lelli_dynamics_2012a,lelli_dynamics_2012b, martinez-delgado_dwarfs_2012,lelli_triggering_2014,ashley_hi_2017}, some authors claim gas accretion as a potential trigger for enhanced star formation in BCDs \citep{lopez-sanchez_intriguing_2012,lelli_triggering_2014,ashley_hi_2017}. Systematic searches for intergalactic \ion{H}{i} clouds in the vicinity of BCDs have not been very successful \citep{taylor_h_1995,taylor_h_1996,taylor_h_1997} apart from searches in denser environments \citep{hoffman_neutral_2003}. However, some studies report of evidence for the presence of companion gas clouds \citep{stil_hi_2002,thuan_h_2004,lopez-sanchez_intriguing_2012} or potential cold gas accretion \citep{lopez-sanchez_intriguing_2012,
lelli_triggering_2014,ashley_hi_2017}. There is hence some evidence of cold accretion or potentially reaccretion after having triggered starburst episodes in dwarf galaxies, but this comes mostly from studies of BCDs. Very few attempts have been made \citep{kamphuis_warp_2011,schmidt_structure_2014} to find evidence of this mechanism in low-mass galaxies and hence the progenitors of accreting starburst galaxies, i.e., dIs with low star formation rate which, before becoming BCDs, show some evidence for infalling gas in their morphology and kinematics.\\
\indent
Theory predicts that if at $z\,=\,0$, the so-called cold accretion mode of accretion from the IGM to the ISM of galaxies still takes place this happens in intermediate and smaller dark matter halos \citep[M$_{\rm vir} < 10^{11}$,][]{Nelson2013,Huang2019}. However, for the smallest of these (V$_{\rm rot} < 40$ km s$^{-1}$) cold gas accretion is likely to be suppressed once more, through the
photo-heating from the UV-background \citep{hoeft_dwarf_2010}. Regardless of the current star formation activity, which is expected to be low in a starburst progenitor system, with gas depletion times larger than the Hubble time, the investigation of neutral hydrogen (\ion{H}{i}) in potentially accreting dwarf systems can hence serve as important input to galaxy formation theory.\\
\indent
This paper presents a study of a potential accretion event of neutral gas with a low SFR, which might later evolve into a starburst.\\
\indent
For the analysis of the Local Volume \ion{H}{i} Survey 
%%%
%%%
\citep[LVHIS,][]{koribalski_local_2018}
%%%
%%%
we fitted all galaxies with the tilted-ring model \citep{Wang2017} by using the automated 3D code {\sc FAT}\footnote{\url{https://github.com/PeterKamphuis/FAT}} \citep{Kamphuis2015}. While most galaxies were properly fitted the galaxy ESO~149-G003 stood out due to a fitted rotation curve (RC) that was close to zero in the inner parts and showed a steep rise in the outer part. For this reason, we did a closer visual inspection of the LVHIS observations which showed there was indeed gas at anomalous velocities in this galaxy, with a velocity gradient perpendicular to the rotating disc. In addition to the kinematically anomalous gas and a slight asymmetry, the data show marginal indications of an extension of the anomalous gas towards a morphologically offset component in extension of the anomalous gas. As laid out above, while the presence of kinematically anomalous gas is not surprising, intergalactic clouds close to dIrr galaxies, which do not form stars at a massive rate, are not regularly observed. The LVHIS results hence motivated further observations with better sensitivity and resolution to confirm the presence of the anomalous gas, to locate the extraplanar component, to estimate its mass, and a first attempt to understand its origin, which is the content of this paper.\\
\indent ESO 149-G003 is an irregular dwarf galaxy at a distance of $7.0\,\mathrm{Mpc}$ \citep{Tully2013} and seen at a high inclination. Table \ref{tab:gal_prop} gives an overview of several galaxy properties. Unfortunately, the original LVHIS data lacked the spatial resolution as well as the sensitivity to determine the nature of this anomalous gas.\\
\begin{table*}
    \centering
    \begin{tabular}{llrlc}
       \multicolumn{5}{c}{ESO 149-G003}\\
       \hline
       \hline
	Parameter & Symbol & Value & Unit & Ref.\\
       \hline
       	Morphological Type & & dI & & 2\\
	Optical centre & RA &$23^{{\rm h}}\, 52^{{\rm m}}\, 2.\!\!^\mathrm{s}8$& & 3\\
	(J2000)& Dec &$-52^{\circ}\,34\arcmin\, 39.\!\!\arcsec7$ & & \\
	Systemic Velocity & $v_{\rm sys}$ & $577.8\,\pm\,0.4$& $\mathrm{km}\,\mathrm{s}^{-1}$ & 1\\              		
 	Distance & $d$ & $7.0$ & $\mathrm{Mpc}$ & 4 \\
	B-band magnitude & $M_\mathrm{B}$ & $-14.\!\!^\mathrm{m}27$ &  & 5\\
	Blue stellar luminosity & $L_\mathrm{B}$ & $7.7\cdot 10^7$ & $L_\odot$ & 5\\
	K-band stellar mass & $M_*(\mathrm{K})$ & $ 5.3\cdot 10^7 $& $M_\odot$ &  1 \\
	Star formation rate & $SFR(\mathrm{H\alpha})$ & $0.006 $ &  $M_\odot\,\mathrm{yr}^{-1}$ & 5 \\
	Star formation rate & $SFR(\mathrm{UV,FIR})$ & $0.010 $ & $M_\odot\,\mathrm{yr}^{-1}$ & 6 \\
	Optical diameter & $D_{25}$ &$131\arcsec\, \pm\, 7\arcsec$&  & 7\\
	Optical diameter & $D_{25}$ &$4.5\, \pm\, 0.3$& $\mathrm{kpc}$& 7\\
	Optical position angle & $PA_\mathrm{opt}$ & $148\deg$ &   & 7\\
	\ion{H}{i} diameter & $D_\ion{H}{i}$ & $173\arcsec\,\pm\, 3\arcsec$ & &1\\
	\ion{H}{i} diameter & $D_\ion{H}{i}$ & $5.9\,\pm\,0.1$ & $\mathrm{kpc}$ & 1\\
	&$D_\ion{H}{i}$/D$_{25}$&$1.32\, \pm\, 0.07$ & &1\\
	Total \ion{H}{i} Mass& $M_\ion{H}{i}$ &$7.1\, \pm\, 0.6\cdot 10^7$& $M_\odot$ &  1\\  
	Rotation velocity & $v_\mathrm{rot}(r_\ion{H}{i})$ &$18.3\,\pm\,1.4$& $\mathrm{km}\,\mathrm{s}^{-1}$ & 1\\
	Dyn. Mass & $M_\mathrm{dyn}(r_\ion{H}{i})$ & $2.3\,\pm\,0.4\cdot 10^8$&$M_\odot$ & 1\\
	\ion{H}{i} line width & $W_{50}$ & $37\,\pm\, 3 $ &
	$\mathrm{km}\,\mathrm{s}^{-1}$ & 1\\
    & $W^\mathrm{c}_{50}$ & $37\,\pm\, 3 $ & $\mathrm{km}\,\mathrm{s}^{-1}$ & 1\\
    & $W_{20}$ & $65\,\pm\, 5 $ & $\mathrm{km}\,\mathrm{s}^{-1}$ & 1\\
    & $W^\mathrm{c}_{20}$ & $67\,\pm\, 5 $ & $\mathrm{km}\,\mathrm{s}^{-1}$ & 1\\
    \hline
    \end{tabular}
     \caption{Fundamental properties of ESO~149-G003. (1) This work, (2) \citet{Laubarts1982}, (3) \citet{Loveday1996}, (4) \citet{Tully2013}, (5) \citet{karachentsev_star_2013}, (6) \citet{wang_local_2017}, (7) \citet{deVaucouleurs1991}.}
    \label{tab:gal_prop}
 \end{table*}
\indent In this paper we present MeerKAT-16 follow up observations of the dwarf galaxy ESO 149-G003 and observations with the optical MeerLICHT telescope, and their subsequent analysis to investigate the anomalous gas component in ESO~149-G003. With a current star formation rate between $0.006 M_\odot\,\mathrm{yr}^{-1}$ and $0.01 M_\odot\,\mathrm{yr}^{-1}$ and a gas mass of $7.1\cdot 10^7\,M_\odot$, the neutral gas depletion time lies between $7\,\mathrm{Gyr}$ and $12\,\mathrm{Gyr}$, close to a Hubble time. ESO~149-G003 has hence a slightly lower gas depletion time compared to the average LSB population \citep{van_zee_evolutionary_2001}, but is still a dIrr galaxy. We hence utilize very early science verification data of two new telescopes to show that this galaxy might exhibit a rather rare event of gas accretion before a possible starburst in ESO~149-G003 sets in.\\
\indent
The paper is structured as follows: $\S$\ref{sect:HI_observations} presents the new \ion{H}{i} observations of ESO 149-G003 including any specifics of the telescope that was still in commissioning and $\S$\ref{sect:HI_DR} presents the data reduction, as well as the optical observations with MeerLICHT and the corresponding data reduction.  
In $\S$\ref{sect:GDaM} we describe the resulting data and a kinematical model based on the radio data. In $\S$\ref{sect:D} we discuss potential physical interpretations of our findings and attempt to interpret them in terms of possible scenarios for the generation of the kinematical structure that we observe. We summarize our results in $\S$\ref{sect:SaO}. In the appendix we show ancillary figures for the interested reader. Throughout the paper we are employing a distance of precisely $7.0\,\mathrm{Mpc}$ \citep{Tully2013}. Distance-dependent quantities from the literature are corrected to this distance.

\section{Observations and data reduction}
\label{sect:observations}
\subsection{MeerKAT observations}
\label{sect:HI_observations}
\begin{table}
    \centering
    \begin{tabular}{@{} lll @{}}
       \multicolumn{2}{c}{Observational Parameters}\\
       \hline
       \hline
	Parameter &  \\
       \hline
    Target &  ESO 149-G003\\
       	Observation Dates & 3 and 4 May, 2018\\
       	Bandpass/Flux Calibrator  & J0137+3309\\
       	Gain Calibrator & J2329-4730\\
       	Time on target & 12.5 h \\
       	Total observation time & 15 h \\
       Frequency Range & $900 - 1670\, \mathrm{MHz}$ \\
       	Central Frequency & $1285\,\mathrm{MHz}$\\
       	Spectral Resolution & 26.123 kHz \\
       	Number of Channels & 32768\\
       	Number of Antennas & 16\\
       	Resulting rms, per channel & $1.0\,\mathrm{mJy}\,\mathrm{beam}^{-1}$\\
       	Spatial resolution (HPBW) & $47.\!\!\arcsec1 \times 32.\!\!\arcsec6$ \\
       	Position angle of synthesized beam & $126.\!\!^\circ5$\\
 \hline
    \end{tabular}
     \caption{The details of the MeerKAT observations used in this study}
    \label{tab:obs_info}
 \end{table}ESO 149-G003 was observed in May 2018 with a MeerKAT-16 array as part of the MeerKAT Early Science Program. Detailed information about the MeerKAT telescope can be found in \citet{jonas_meerkat_2016} and \citet{camilo_revival_2018}. The specifics of the observations are presented in Table ~\ref{tab:obs_info}. The observations were split into two days, with approximately 6 hours of on-target observations on each day at different times to maximise the $u-v$ coverage. \newline
\indent The data were taken with the L-band receiver of MeerKAT which extends from $900-1670\,\mathrm{MHz}$. For the bandpass and flux calibration we observed 3C48 for 5 minutes every hour except the first three hours on each day. Additionally, we observed the gain calibrators J2357-5311 and J2329-4730 for 2 minutes every 20-25 minutes. Only 16 antennas of the array were used; with the longest baselines being $\sim 3$ km. The observations were conducted using three different subarrays with different antenna configurations, necessitating separate calibration for each, which we describe in the next section.
\subsection{Reduction of radio data}\label{sect:HI_DR}
For the purposes of this work we have chosen a frequency window of $40$ MHz centred around 1400 MHz, primarily to keep the reduction time, processing and storage requirements to a minimum. The frequency resolution of $26.12$ kHz gives a total of $1534$ channels across our chosen frequency window. \newline
\indent The data were reduced with
the Containerized Automated Radio Astronomy Calibration (CARACal)  pipeline\footnote{\url{https://caracal.readthedocs.io}} developed at SARAO and INAF \citep{Jozsa2020}.
It is a containerised pipeline for the reduction of 
interferometric spectral line  and continuum data, based on the {\sc Python} scripting framework {\sc Stimela}\footnote{\url{https://github.com/ratt-ru/Stimela
}}, allowing users to run
several radio interferometry software packages in
a same script. 
\begin{figure}
\includegraphics[width=\columnwidth]{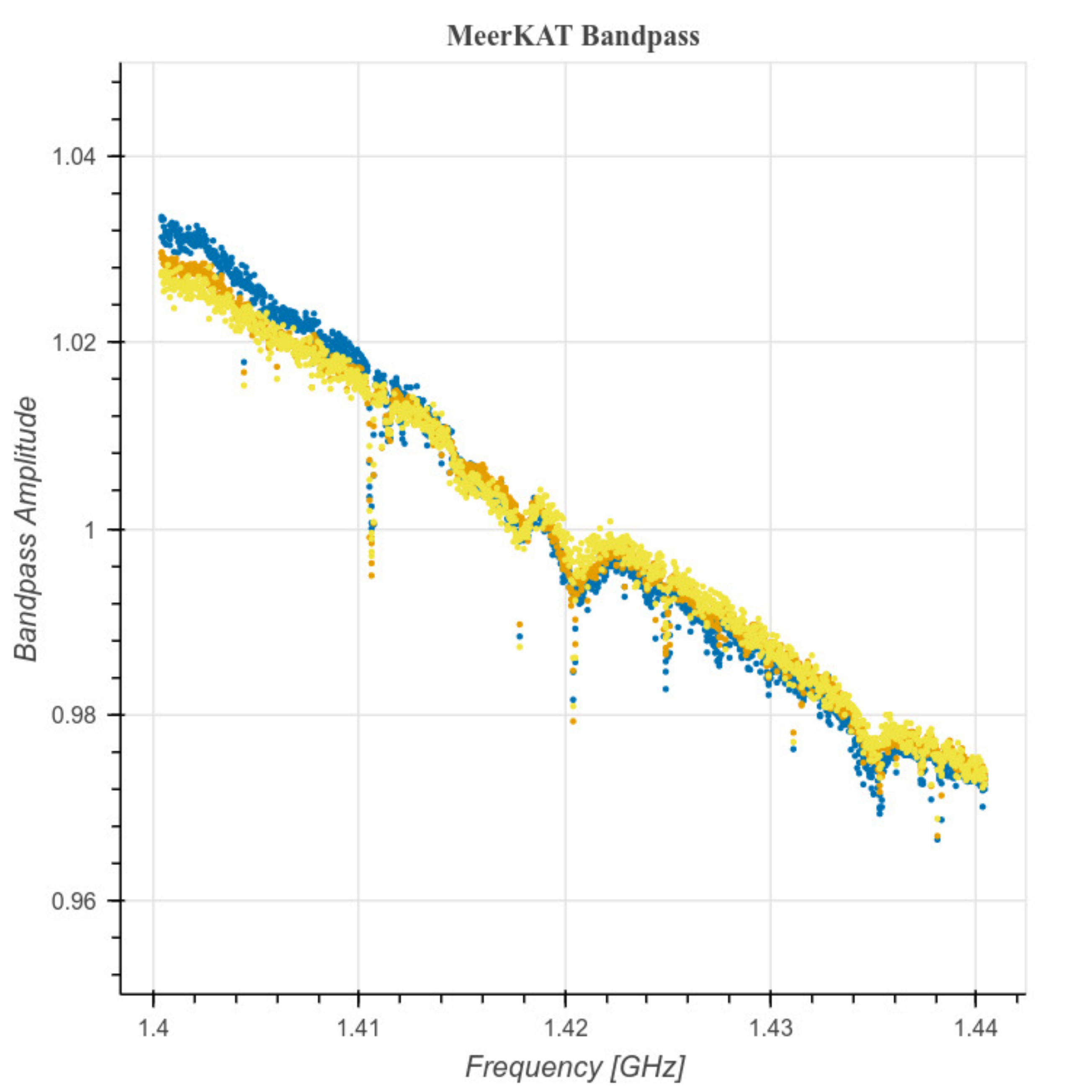}
\caption{The estimated bandpass for MeerKAT-16, different colours represent different bandpass solutions corresponding to each subarray (to each antenna configuration observed with at different times, see $\S$\ref{ccp}).}
 \label{fig:bandpass}
\end{figure}
\subsubsection{RFI mitigation}
As stated in $\S$~\ref{sect:observations}, the observations were done with three different sub-arrays resulting in three measurement sets. A similar data reduction procedure was conducted for all measurement sets. Each of the measurement sets was put through initial flagging of radio frequency interference (RFI) of the calibrator fields only, carried out with CASA Flagdata task as well as AOFlagger \citep{Offringa2010}. To avoid flagging unaffected data, we wrote a custom flagging strategy for AOFlagger. This strategy uses lower starting sensitivity for the sumthreshold algorithm than the standard AOFlagger strategy, as well as involves more iterations and the use of dilation of flags in time-frequency plane to comprehensively remove the RFI. Additionally, we flag the \ion{H}{i} emission from Milky Way manually. \newline
\subsubsection{Cross-Calibration Process}\label{ccp}
A standard cross-calibration using CASA \citep{McMullin2007}, including delay calibration, bandpass calibration, and gain calibration, was applied for all three measurement sets taken (using three separate antenna configurations or subarrays). The Perley-Butler 2010 standard was used to set the flux scale \citep{PerleyButler13}. It should be noted that the data show "spikes" of unknown origin, which in turn are seen in the estimated bandpass as dips. These dips may potentially hinder detection and/or characterisation of spectral lines, in case of imprecise or insufficient bandpass corrections. Fortunately, we did not see any significant dips in the bandpass in the frequencies corresponding to the \ion{H}{i} emission from ESO 149-G003 (20 channels in the frequency range $1417.475\,\mathrm{MHz} - 1417.971\,\mathrm{MHz}$). We show a plot of the estimated bandpass in Fig.~\ref{fig:bandpass}, where different colours denote the separate measurement sets. From this figure we can see that the characteristics of the "dips" seems stable with time, since the solutions shown are calculated for separate observations which were conducted at a different time. The target data was subsequently flagged after the cross-calibration. 
\subsubsection{Self-Calibration Process}
The cross-calibrated data were put through four iterations of  self-calibration, each consisting of imaging and calibration. For the imaging we used the WSClean imager \citep{Offringa2014}. For self-calibration, we used the CUBICAL software \citep{Kenyon2018}. Each imaging cycle was followed by a calibration step, using a sky model generated from the clean components. The deconvolution, to this end, was made with successively lower thresholds, using the auto-masking and auto-thresholding features of WSClean. The calibration loops contained a single round of phase calibration and three rounds of amplitude-phase calibration. With this self-calibration process, we were able to achieve an rms of $0.1\, \mathrm{mJy}\,\mathrm{beam}^{-1}$ in the continuum image.
During the calibration process, we used a binning of three channels while imaging the data. Since the model is formed from the clean components, this may cause a step-function in the model visibilities with frequency, with the step size being the size of one of the frequency bins used for imaging. This can cause issues in the process of continuum subtraction, since the data itself has a much finer resolution. To avoid this, in the last imaging round, we imaged the data at the finest possible resolution (while using joint deconvolution). 

Post-calibration, we imaged the data such that the output data cube (and thus the model data) contains the same number of channels as data - this can be contrasted with imaging during calibration which has three channels in the output images, while jointly deconvolving the channels together. This was done to avoid the presence of a "step-function" in the model data which would affect the continuum subtraction. The rms per channel in the resulting data cube is $1\,\mathrm{mJy}\,\mathrm{beam}^{-1}$ over a channel of $26.1\,\mathrm{kHz}$.
\newline
\subsubsection{Continuum Subtraction and Imaging}
\begin{figure*}
 \includegraphics[width=\textwidth]{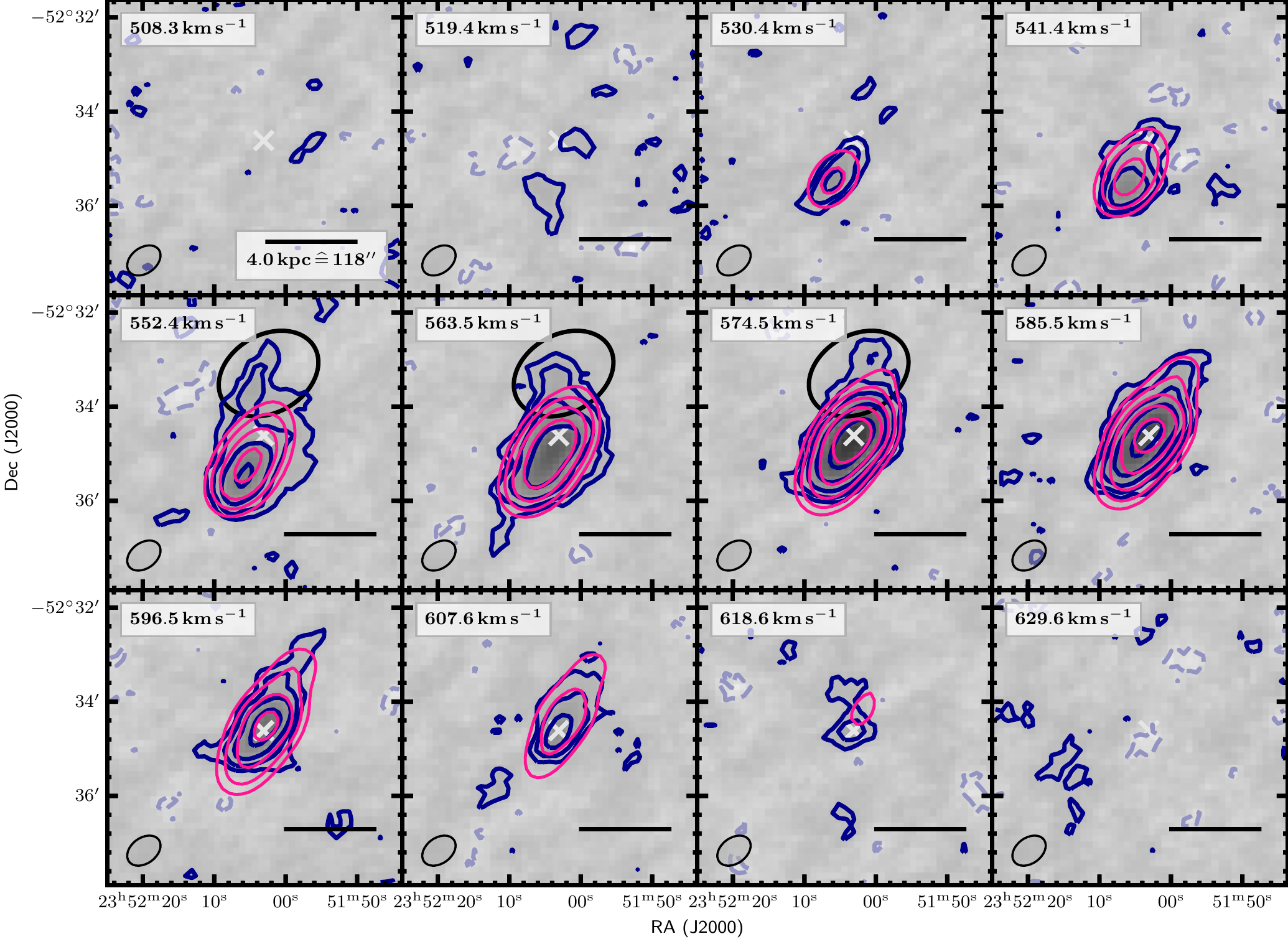}
 \caption{MeerKAT \ion{H}{i} data cube, omitting every second channel. Contours denote the $-2,2,4,8,16,32\,-\,\sigma_\mathrm{rms}$-levels, where $\sigma_\mathrm{rms}\,=\,1\,\mathrm{mJy}\,\mathrm{beam}^{-1}$ . Blue: the observed data cube. Pink: the {\sc TiRiFiC} model. Dashed lines represent negative intensities. The white cross represents the kinematical centre of the model, the ellipse to the lower left the syntesized beam ($HPWB$), the black ellipse to the top highlights kinematically anomalous gas. A full data cube can be found in Figs.~\ref{fig:ESO149-G003_data_cube_24_obs} and \ref{fig:ESO149-G003_data_cube_24} in the appendix.}
 \label{fig:ESO149-G003_data_cube_12}
\end{figure*}

\begin{figure*}
 \includegraphics[width=\textwidth]{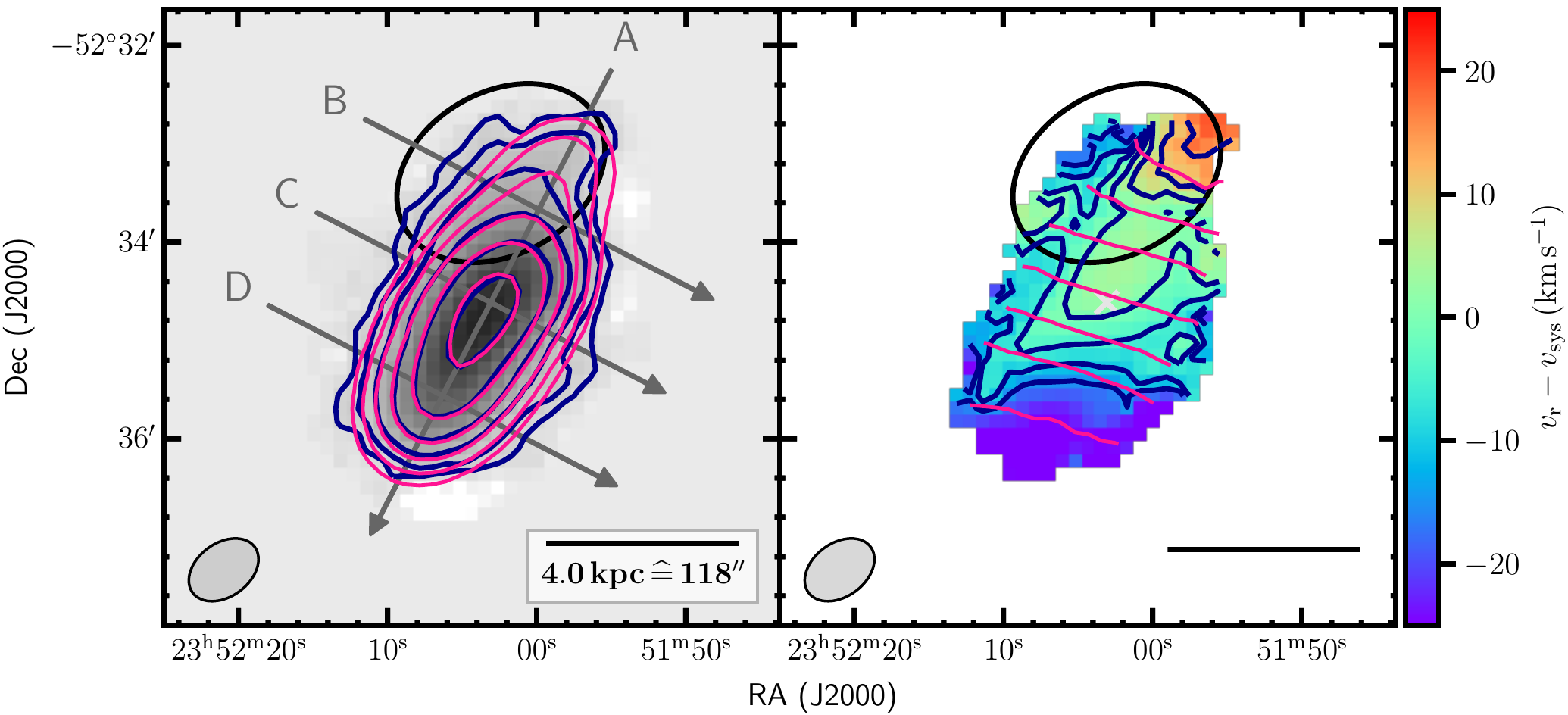}
 \caption{MeerKAT \ion{H}{i} total-intensity map and moment-1 velocity field, as derived from the data cube shown in Figs.~\ref{fig:ESO149-G003_data_cube_12}, \ref{fig:ESO149-G003_data_cube_24_obs}, and \ref{fig:ESO149-G003_data_cube_24}. Blue: derived from the observation. Pink: derived from the {\sc TiRiFiC} model. The white cross or the intersection between arrows A and C represents the kinematical centre of the model, the ellipse to the lower left the syntesized beam ($HPWB$), the black ellipse to the top highlights the position of kinematically anomalous gas. Left: Total-intensity map. Contours denote the $0.25,0.5,1,2,4,8\,$-$\,M_\odot\,\mathrm{pc}^{-2}$-levels. Arrows indicate the positions of slices along which the position-velocity diagrams in Fig.~\ref{fig:ESO149-G003_PV-diagrams} were taken. Right: velocity field. Contours are isovelocity contours. They denote the $-15, -10, -5, 0, 5, 10,15\,\mathrm{km}\,\mathrm{s}^{-1}$-levels relative to the systemic velocity $v_\mathrm{sys}\,=\,578\,\mathrm{km}\,\mathrm{s}^{-1}$. A version showing only the data derived from the observed data cube can be found in the appendix (Figs.~\ref{fig:ESO149-G003_mom0_mom1_vf_obs}).}
\label{fig:ESO149-G003_mom0_mom1_vf}
\end{figure*}

The continuum subtraction process was twofold; firstly by subtracting the continuum model from the calibrated data sets and secondly through the CASA task {\sc uvlin} to subtract any residual continuum, if present. Each channel of the continuum-subtracted data was then imaged individually with WSclean and combined into a data cube of \ion{H}{i} emission in ESO 149-G003.\\
\indent The imaging  was done in an iterative manner with three rounds of imaging, in which the first two imaging rounds were used to obtain successively better deconvolution masks. For effectively imaging the diffuse emission associated with the \ion{H}{i} source, we carried out the imaging at a Briggs weighting of 2 ("Natural Weighting"), while adding a further tapering which gave a final resolution of $HPBW\,=\,47.\!\!\arcsec1 \times 32.\!\!\arcsec6$. In Fig.~\ref{fig:ESO149-G003_data_cube_12} we present selected channels from the resulting data cube (omitting every second channel), while Figs.~\ref{fig:ESO149-G003_data_cube_24_obs} and \ref{fig:ESO149-G003_data_cube_24} in the Appendix show the full data cube.\\
%%%%
%%%%
\indent To construct the \ion{H}{i} total-intensity map, we employed several {\sc Miriad}\footnote{We were using the \textit{Carma} version of {\sc Miriad} (\url{http://admit.astro.umd.edu/miriad/}), and several {\sc Miriad} tasks adjusted by ourselves.} tasks. We created a mask by convolving the cube to a resolution of $60^{\prime\prime}$ and consecutive clipping at a level of $3\,\rm{mJy}\,\rm{beam}^{-1}$, corresponding to $2.5\,\sigma_{\rm rms}$ in the convolved cube. The resulting mask was further processed using the task {\sc mafia} to contain only fields above the clip level which overlapped in two consecutive channels. With the chosen clip level only one masked field remained containing the \ion{H}{i} emission of ESO~149-G003. We then used our kinematic galaxy model as described in $\S$\ref{sect:GDaM} to widen the mask. We convolved the model data cube to a resolution of $60^{\prime\prime}$ and added the mask resulting from clipping the convolved model data cube at a level of $0.9\,\rm{mJy}\,\rm{beam}^{-1}$ to the existing one. We then calculated a moment-0 total-intensity map and a moment-1 map using the combined mask. The result is shown in Figs.~\ref{fig:ESO149-G003_mom0_mom1_vf} and \ref{fig:ESO149-G003_mom0_mom1_vf_obs}.\\
\indent
We calculated the \ion{H}{i} spectrum of ESO~149-G003 (Fig.~\ref{fig:ESO149-G003_spectrum}) by applying the same mask. The total flux of $F_\ion{H}{i}\,=\,6.1\,\pm\,0.5\,\mathrm{Jy}\,\mathrm{km}\,\mathrm{s}^{-1}$, the \ion{H}{i} mass of $M_\ion{H}{i}\,=\,7.1\, \pm\, 0.6\, M_\odot$, and the spectral width at 50\% and 20\% of the peak flux density (corrected for an inclination of $79\deg$, see table~\ref{tab:model_prop}) $W_{50}\,=\,37\,\pm\, 3 \,\mathrm{km}\,\mathrm{s}^{-1}$ and $W_{20}\,=\,67\,\pm\, 5 \,\mathrm{km}\,\mathrm{s}^{-1}$ were derived from this spectrum, with the errors estimated from our tilted-ring model (see $\S$\ref{sect:trm}) by creating a model bearing a maximal and a minimal velocity width given the limits of the model errors. Given a good agreement between modelled and original spectrum we interpret the difference as the true error for above quantities (see also Table~\ref{tab:gal_prop}). The total flux of $F_\ion{H}{i}\,=\,6.1\,\pm\,0.5\,\mathrm{Jy}\,\mathrm{km}\,\mathrm{s}^{-1}$ is in good agreement with the total \ion{H}{i} flux of $6.9\,\pm\,1.6\,\mathrm{Jy}\,\mathrm{km}\,\mathrm{s}^{-1}$ as derived in the HIPASS bright galaxy catalogue \citep{Koribalski2004} and with $F_\ion{H}{i}\,=\,7.2\,\mathrm{Jy}\,\mathrm{km}\,\mathrm{s}^{-1}$ from the LVHIS catalogue 
%%%
%%%
\citep{koribalski_local_2018}.
%%%
%%%
Also $W_{50}$ and $W_{20}$ are in reasonable agreement ($W_{50}^\mathrm{HIPASS}\,=\,39\,\pm\, 10\,\mathrm{km}\,\mathrm{s}^{-1}$, $W_{20}^\mathrm{HIPASS}\,=\,70\,\pm\, 15\,\mathrm{km}\,\mathrm{s}^{-1}$).
%%%
%%%
%%%%
\subsection{MeerLICHT observations} \label{sect:meerlicht_observations_and_dr} 
\begin{figure*}
 \includegraphics[width=\textwidth]{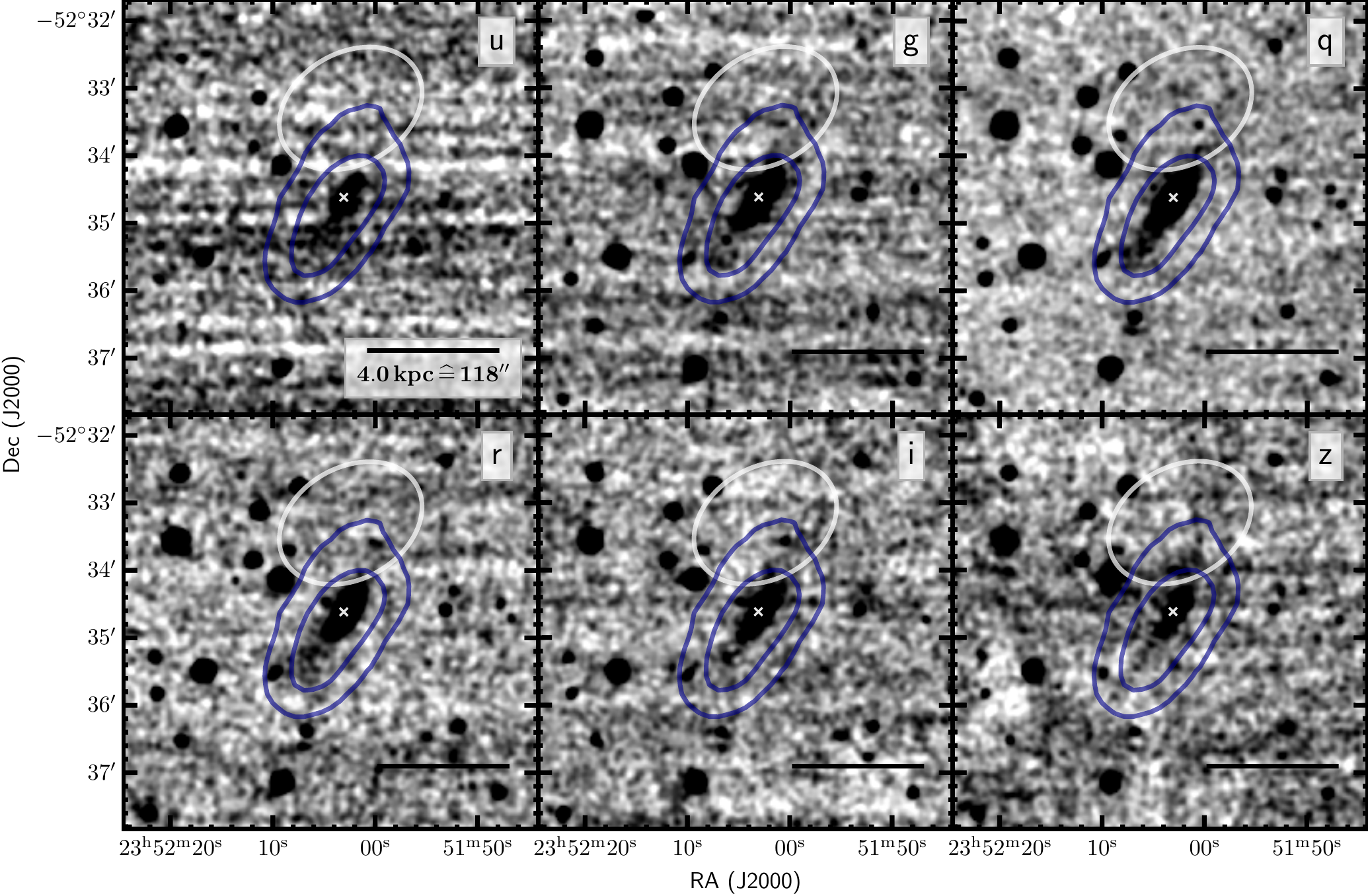}
 \caption{MeerLICHT observations of ESO~149-G003, convolved with a 2D-Gaussian of $FWHM\,=\, 6^{\prime\prime}$ in each direction (RA and Dec). Contours denote the $1,4\,$-$\,M_\odot\,\mathrm{pc}^{-2}$-levels of the observed total-intensity map shown in the left panel of  Figs.~\ref{fig:ESO149-G003_mom0_mom1_vf} and \ref{fig:ESO149-G003_mom0_mom1_vf_obs}. The white ellipse to the top highlights kinematically anomalous gas. The white cross represents the kinematical centre of the kinematical model. The full-resolution maps can be found in Fig.~\ref{fig:ESO149-G003_optical_meerlicht} in the appendix.}
 \label{fig:ESO149-G003_optical_meerlicht_con_6}
\end{figure*}

To examine the environment of ESO 149-G003, as seen at optical bands, we used the MeerLICHT telescope\footnote{MeerLICHT is an optical wide-field telescope, located at the Sutherland station of the South African Astronomical Observatory, South Africa. It has a 0.65 m primary mirror allowing for a field-of-view of 2.7 square degrees, sampled at 0.56\arcsec/pixel} \citep{bloemen_meerlicht_2016}. ESO~149-G003 was observed in the bands u,g,q,r,i and z. Here we give a very brief summary of the optical data observations and reductions.\newline
\indent In each band the target was observed for a total of 5 minutes in single exposures of 1 minute. The image pre-processing consisted of gain-, crosstalk-, overscan-, master flat-, cosmic ray-, and satellite trail corrections.\newline
\indent To obtain the astrometric solution, a mask for bad pixels and edge pixels on the CCD, saturated pixels and pixels around saturated pixels as well as satellite trails and cosmic rays was created. The 1000 brightest stars which do not fall on the masked pixels were then selected (the source finding was done with SExtractor, \citealt{Bertin1996}, with a detection threshold of $5\sigma$). These stars were then compared with the the MeerLICHT calibration catalog, based on {\it Gaia} DR2 data \citep{gaia_gaia_2016,gaia_gaia_2018}. The comparison between the stars detected in the images and the stars in the MeerLICHT calibration catalog was used to generate the astrometric solution for the images. \newline
\indent
For our analysis, we stacked (using median stacking) the individual images in each band to create a single image per band. These images are shown in Fig.~\ref{fig:ESO149-G003_optical_meerlicht_con_6}.  
%%%%
%%%%
\section{Gas distribution and modelling} \label{sect:GDaM}
\subsection{Data inspection}\label{sect:GDaM_data_inspection}
\ion{H}{i} associated with ESO~149-G003 can be seen in the data cube (Figs.~\ref{fig:ESO149-G003_data_cube_12}, \ref{fig:ESO149-G003_data_cube_24_obs}, and \ref{fig:ESO149-G003_data_cube_24}) in emission in $18-20$ channels of $5.5\,\mathrm{km}\,\mathrm{s}^{-1}$ width. Figures~\ref{fig:ESO149-G003_mom0_mom1_vf} and \ref{fig:ESO149-G003_mom0_mom1_vf_obs} show the total \ion{H}{i}-intensity distribution (left) and the velocity field of ESO~149-G003 (right). The total intensity is elongated and centrally concentrated, which is indicative for a filled, flat structure, and the isovelocity contours show a gradient along the \ion{H}{i} body. Both the morphology and the kinematics of the galaxy are hence symmetric to zeroth order, indicating a disc-like structure. At the same time significant deviations from cylindrical symmetry in morphology and strong indications of non-circular motions are visible. Two features stand out: along the major axis the \ion{H}{i} distribution appears more elongated towards the South-East and more devoid of gas on the North-West, indicating a global morphological asymmetry; in addition a component to the North can be observed, which seems to be more localized and stands out with respect to the main gaseous body. In the data cubes (Figs.~\ref{fig:ESO149-G003_data_cube_12} and \ref{fig:ESO149-G003_data_cube_24}) and the moment maps \ref{fig:ESO149-G003_mom0_mom1_vf} we highlight the presence of the Northern extra component with an ellipse. The gas appears to be a localized excess to the North of the galaxy, and is hence morphologically distinct extraplanar gas, albeit in projection still connected to the main gaseous body (spatially and kinematically). As the isovelocity contours are not coherently oriented symmetrically with respect to the major axis, the velocity field (Figs.~\ref{fig:ESO149-G003_mom0_mom1_vf} and \ref{fig:ESO149-G003_mom0_mom1_vf_obs}, right panel) shows clear indications of non-circular motions of the gas as an additional feature. To some degree the noncircular kinematics seems to be systematic for the whole galaxy, as is indicated by isovelocity contours coherently being tilted with respect to the morphological (and kinematical) major axis of the \ion{H}{i} disc in the galaxy. The contours are, however, not completely symmetric with respect to the minor axis and seem to become largely irregular in the outskirts, most obvious again in the Northern region highlighted by the ellipses.\\ 
\indent
To confirm that this extra-component is not an artifact from the data reduction we compared our results with the publicly available data cube from LVHIS 
%%%
%%%
\citep{koribalski_local_2018},
%%%
%%%
where it appears just above the noise level. In addition the continuum map does not show any strong source at the position of the Northern extension, such that we exclude that the feature might be due to a failed continuum subtraction.\\
%%%%
\begin{figure}
 \includegraphics[width=\columnwidth]{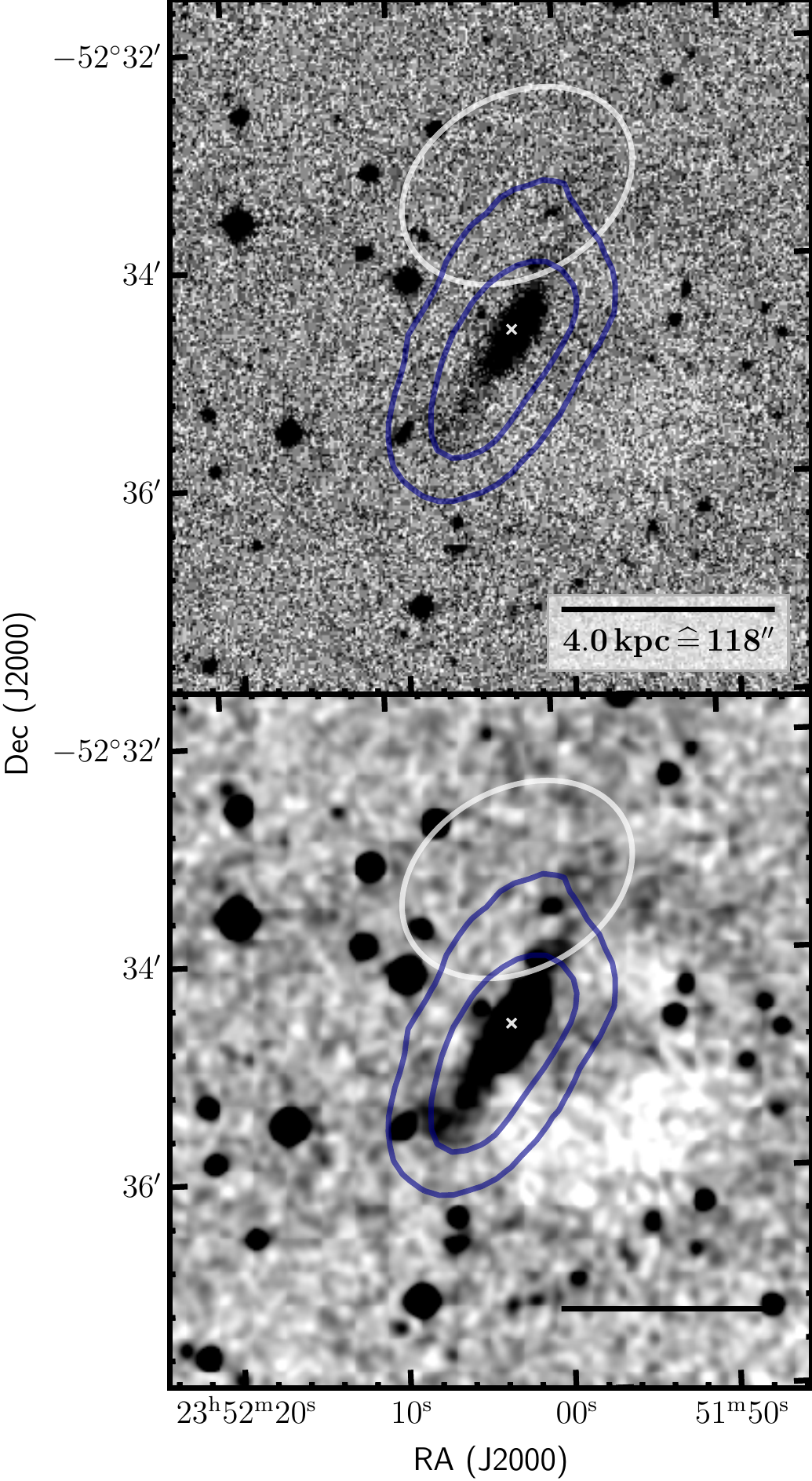}
 \caption{DSS (red, IIIa-F) images of ESO~149-G003. Blue contours denote the $1,4\,$-$\,M_\odot\,\mathrm{pc}^{-2}$-levels of the observed total-intensity map shown in the left panel of  Figs.~\ref{fig:ESO149-G003_mom0_mom1_vf} and \ref{fig:ESO149-G003_mom0_mom1_vf_obs}. The white ellipse to the top in each image highlights the position of kinematically anomalous gas. The white cross represents the kinematical centre of the model. Top: full resolution. Bottom: image in top panel, convolved with a 2D-Gaussian of  $FWHM\,=\, 6^{\prime\prime}$ in each direction.}
\label{fig:ESO149-G003_optical_dss_r_norm_con_6}
\end{figure}
%%%%
\indent Comparing the \ion{H}{i} data to the MeerLICHT images (Figs.~\ref{fig:ESO149-G003_optical_meerlicht_con_6} and \ref{fig:ESO149-G003_optical_meerlicht}) and the publicly available image from the Digitized Sky Survey (DSS)
(Fig.~\ref{fig:ESO149-G003_optical_dss_r_norm_con_6}), we confirm our finding that the \ion{H}{i} distribution is asymmetric and is elongated towards one side (the South-East, approaching) more than into the other (North-West, receding), in particular when orienting on the optical galaxy centre (see Table~\ref{tab:gal_prop} close to the white crosses, see $\S$\ref{sect:kin}), which denote the kinematical centre. We inspected the optical images at native resolution as well as after convolving them with a 2D-Gaussian with a $FWHM$ of $6\arcsec$ in all directions. The elongation to the South-East seems also to be reflected in the stellar distribution, which (after enhancing through convolution) shows a similar asymmetry in the surface density. To highlight this we overlaid \ion{H}{i} contours on the optical images (Figs.~\ref{fig:ESO149-G003_optical_meerlicht_con_6}, \ref{fig:ESO149-G003_optical_dss_r_norm_con_6}, and \ref{fig:ESO149-G003_optical_meerlicht}). In particular the asymmetric elongation to the South-East at the $4\,M_\odot\,\mathrm{pc}^{-2}$-level is also reflected in the optical images. However, we could not find any obvious optical counterpart to the Northern extension, as highlighted by the ellipses in the same images, drawn at the position of the Northern extension. An inspection of an R-band image in \citet{ryan-weber_intergalactic_2004} with a $5-\sigma$ limiting surface brightness of $\mu_\mathrm{R}^\mathrm{AB}\,=\,24.\!\!^\mathrm{m}9\,\mathrm{arcsec}^{-2}$ bears the same results.
\subsection{Tilted-ring modelling}\label{sect:trm}
\begin{figure}
 \includegraphics[width=\columnwidth]{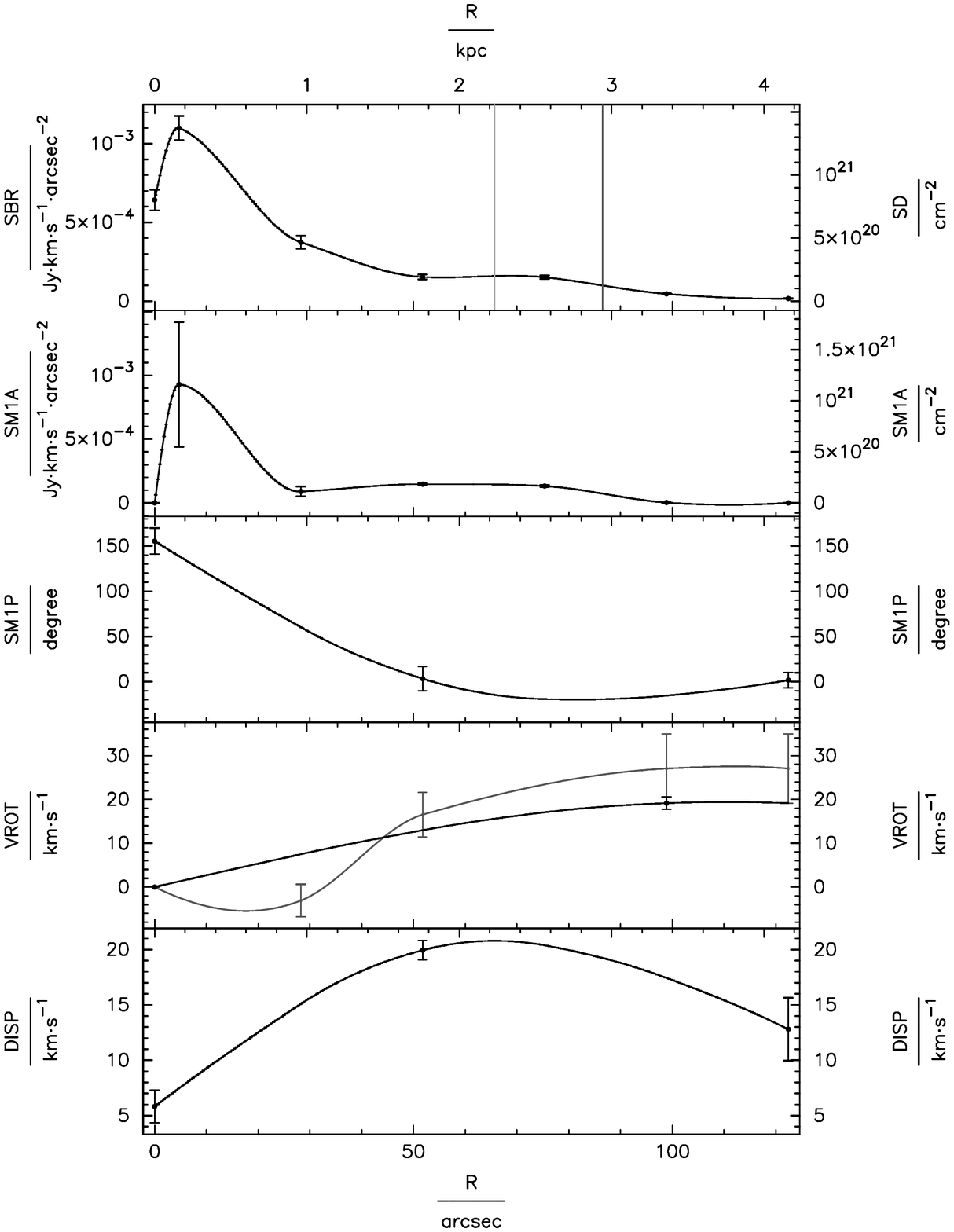}
 \caption{Final {\sc TiRiFiC} model of the \ion{H}{i} disc of ESO~149-G003. SBR/SD: surface brightness or surface column density. The two lines denote $r_{25}$ and $r_\ion{H}{i}$ (at the larger radius). SM1A: Amplitude of a harmonic distortion in surface brightness/surface column density. SM1P: phase of a harmonic distortion in surface brightness/surface column density. VROT: rotation velocity. The grey line is the result of a fit, in which the rotation curve was left to vary on the nodes shown as dots. DISP: dispersion.}
\label{fig:ESO149-G003_modelpars}
\end{figure}

We used the implementations in the software suites {\sc TiRiFiC}\footnote{\url{http://gigjozsa.github.io/tirific/}} and {\sc FAT}~\citep{Jozsa2007,Kamphuis2015} to fit a tilted-ring model to the data. The goal was to derive a simple kinematic model to contrast the observations with and to highlight the differences and peculiarities of the galaxy morphology and kinematics.\\
\indent
The radially invariant parameters of our final model are listed in Table~\ref{tab:model_prop}, while the radially varying parameters are shown in Fig.~\ref{fig:ESO149-G003_modelpars}.\\
%%%%
\begin{table}
    \centering
 \begin{small}
  
    \begin{tabular}{llrl}
       \multicolumn{4}{c}{ESO 149-G003: TiRiFiC radially invariant parameters}\\
       \hline
       \hline
	Parameter & Symbol & Value & Unit\\
       \hline
    Model centre & RA & $23^\mathrm{h}\, 52^\mathrm{ m}\,3.\!\!^\mathrm{s}1 \, \pm\, 0.\!\!\arcsec5$&\\
	(J2000)& Dec & $-52^{\circ}\,34\arcmin\, 37.\!\!\arcsec0\, \pm \, 1.\!\!\arcsec7$ &  \\
	Systemic Velocity & $v_{\rm sys}$ & $577.8\,\pm\,0.4$ & $\mathrm{km}\,\mathrm{s}^{-1}$ \\
	Thickness & $z_0$ & $13\arcsec\,\pm\,4\arcsec$ & \\
	Inclination & $i$ & $79^\circ\,\pm\,7^\circ$ &\\
	Position angle & $pa$ & $ 332.\!\!^\circ6\,\pm\,0.\!\!^\circ7$ & \\
    \hline
    \end{tabular}
    \end{small}
     \caption{Tilted-ring parameters not varying with radius}
    \label{tab:model_prop}
 \end{table}
\indent The adopted model is a flat disc, for which the surface brightness with a harmonic distortion of first order (see below), the rotation velocity, and the dispersion vary with radius (Fig.~\ref{fig:ESO149-G003_modelpars}). All other parameters, the centre and the systemic velocity, the disc thickness, and inclination and position angle do not vary with radius (Table~\ref{tab:model_prop}).
\\
\indent
The final model shows three peculiarities. 
\\
\indent
Firstly, as indicated, to reproduce the asymmetric morphology of the disc we implemented a harmonic distortion in surface brightness (a sinusoidal variation of the surface brightness with a given amplitude $SM1A$ and a phase $SM1P$), which entered as a radially dependent parameter in our final model.
Figure~\ref{fig:ESO149-G003_topview_mom0} in the appendix shows a face-on view of the \ion{H}{i} model, demonstrating that a significant morphological asymmetry had to be introduced, which is also evident from comparing the numerical values of the surface brightness ($SBR$) and the amplitude of the first order harmonic ($SM1A$) in Fig.~\ref{fig:ESO149-G003_modelpars}.\\
\indent
Secondly, we were able to fit a meaningful rotation curve calculating the velocity at different radii via a spline interpolation between three points, the origin (a fixed velocity of $0\,\mathrm{km}\,\mathrm{s}^{-1}$ at radius 0), and two points with identical velocity at larger radii. We hence forced the rotation curve to approach a flat part at the largest radii. This velocity (the rotational amplitude) was consecutively the only variable parameter determining the rotation curve. As soon as more degrees of freedom were included to describe the rotation velocity, the model converged to an unphysical solution (see grey line for $v_\mathrm{rot}$ in Fig.~\ref{fig:ESO149-G003_modelpars}). \\
\indent
Thirdly, the dispersion had to be varied radially and exceeded the rotation velocity at several radii.
\\
\indent
The final values and errors of our model parameters were estimated using a bootstrap method: after converging to a preliminary final model we shifted the single nodes of the model multiple times by a random value and re-started the fitting process. After 20 such fitting processes we calculated the means and the standard deviations for each of the model parameters and assumed the to be the final model parameters and their errors.\\
\indent The \ion{H}{i}-diameter $D_\ion{H}{i}$ as provided in Table~\ref{tab:gal_prop} was derived as the radius at which the surface-mass density of the model reaches $1\,M_\odot \,\mathrm{pc}^{-2}$, and the specific \ion{H}{i} rotation velocity $v_\mathrm{rot}(r_\ion{H}{i})$ as the velocity at the corresponding radius $r_\ion{H}{i}$. The dynamical mass $M_\mathrm{dyn}(r_\ion{H}{i})$ shown in Table~\ref{tab:gal_prop} is the hypothetical gravitational mass of a spherically symmetric mass distribution enclosed at radius $r_\ion{H}{i}$, causing a test particle at that radius to be on a circular orbit with a velocity of $v_\mathrm{rot}(r_\ion{H}{i})$. With $2.3\pm0.4\cdot 10^8\,M_\odot$ it amounts to approximately 1.5 times the sum of $M_\ion{H}{i}$ and $M_*$.
%%%
\subsection{Comparison of TiRiFiC model and data}\label{sect:GDaM_comparison}
\begin{figure*}
 \includegraphics[width=0.45\textwidth]{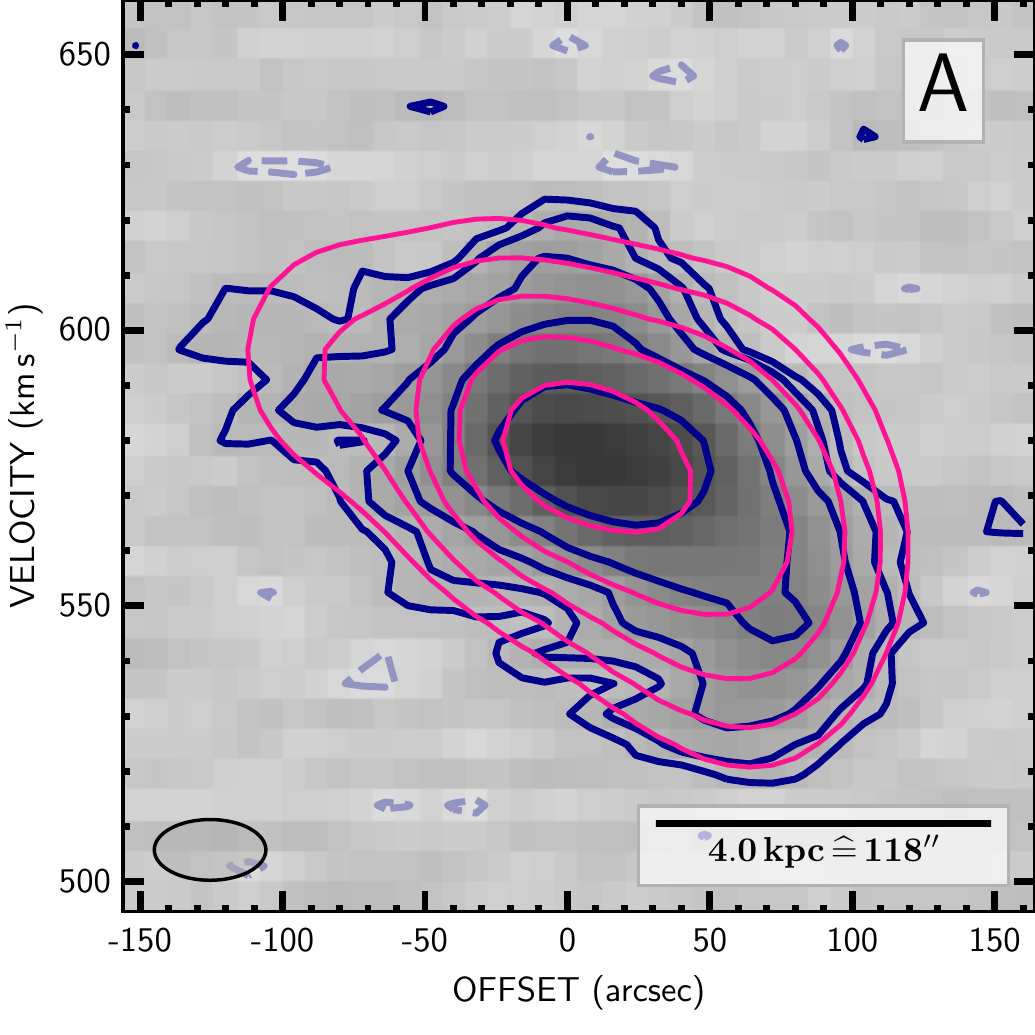}
 \includegraphics[width=0.45\textwidth]{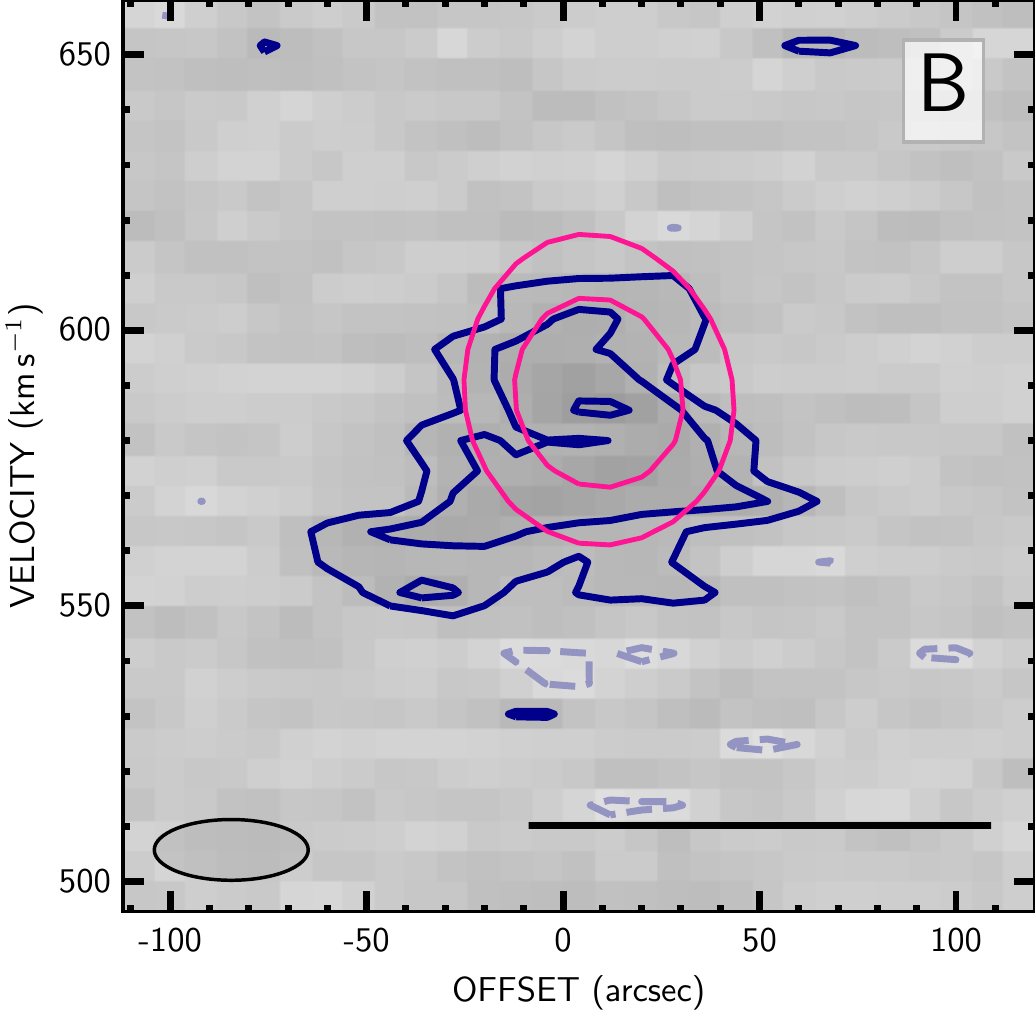}
 \includegraphics[width=0.45\textwidth]{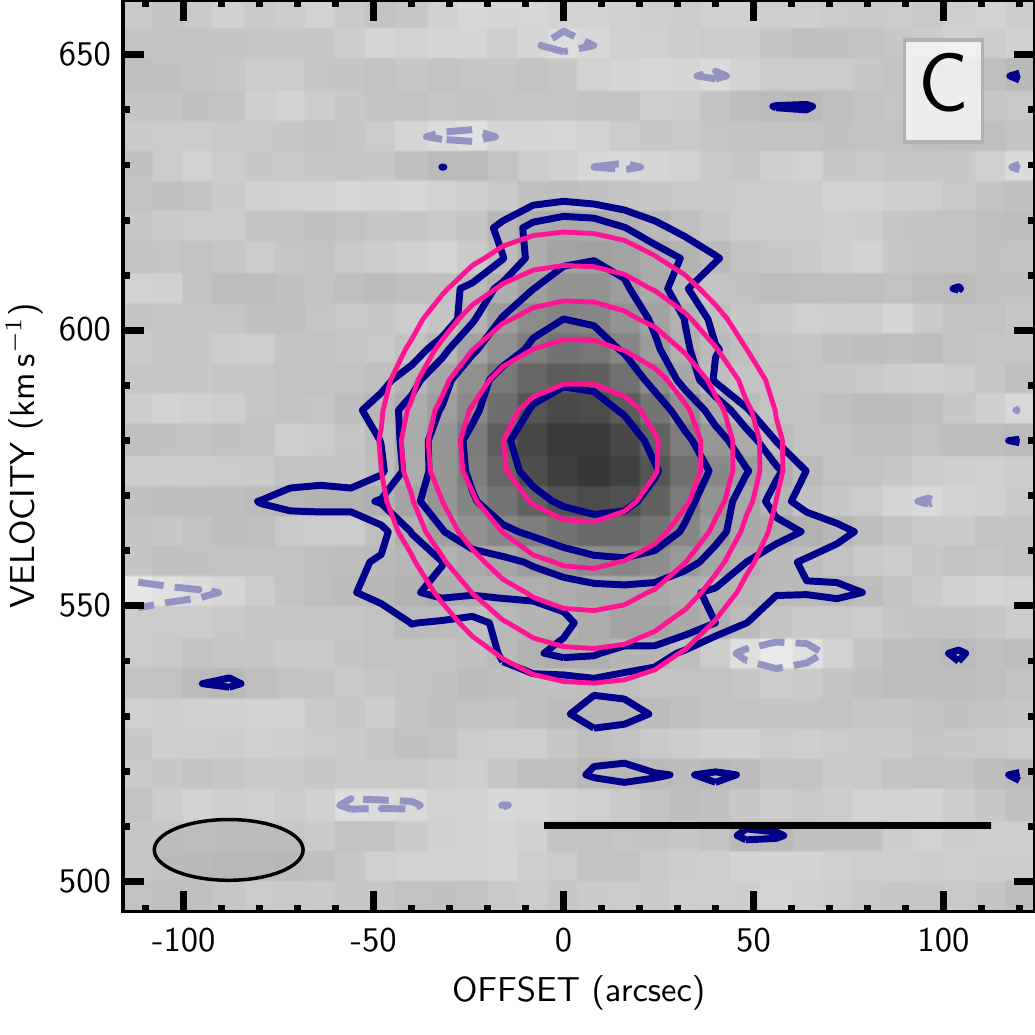}
 \includegraphics[width=0.45\textwidth]{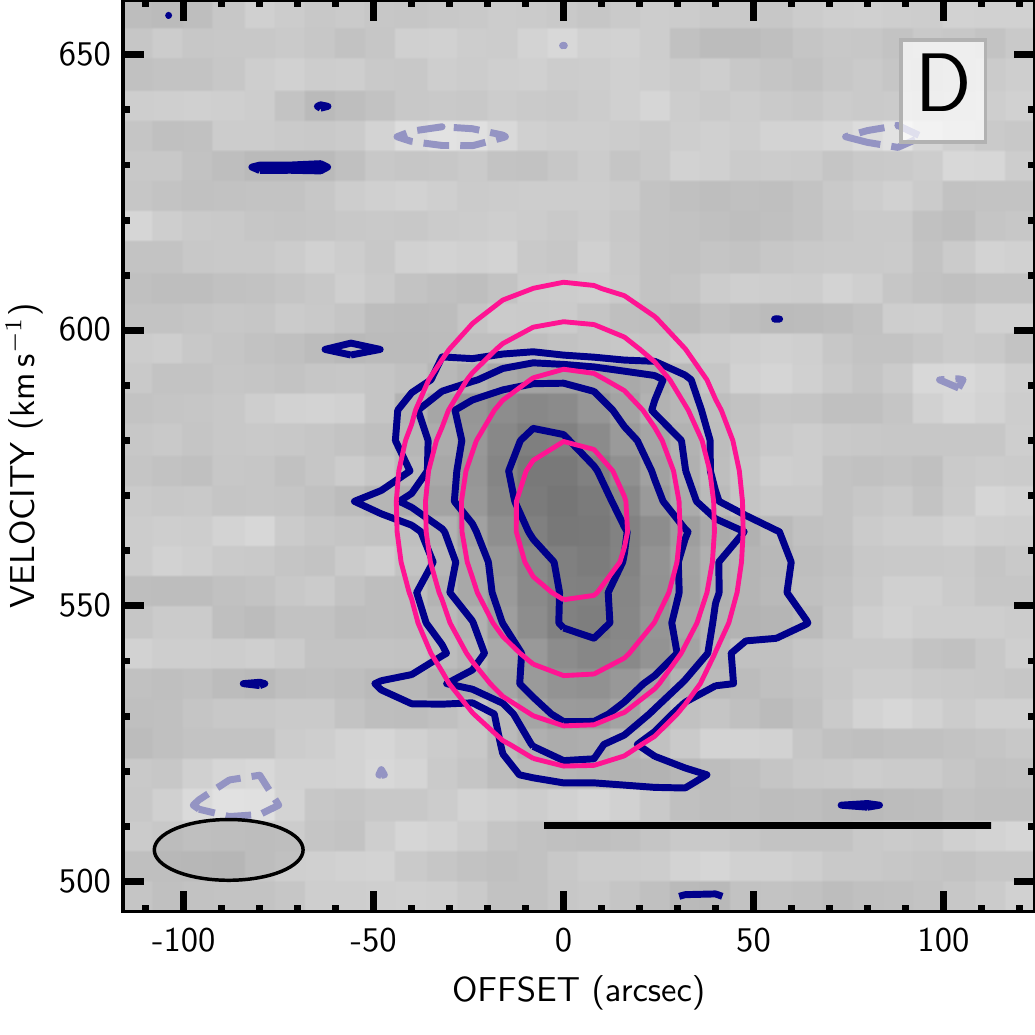}
 \caption{Position-velocity diagrams taken along slices shown in Fig.~\ref{fig:ESO149-G003_mom0_mom1_vf}. Contours denote the $-2,2,4,8,16,32\,-\,\sigma_\mathrm{rms}$-levels, where $\sigma_\mathrm{rms}\,=\,1\,\mathrm{mJy}\,\mathrm{beam}^{-1}$. Blue: the observed data cube. Pink: the {\sc TiRiFiC} model. Dashed lines represent negative intensities. The ellipse represents the velocity (2 channels) and the spatial ($\sqrt{HPWB_\mathrm{maj}*HPWB_\mathrm{min}}$, where $HPWB_\mathrm{maj}$ and $HPWB_\mathrm{min}$ are the major and minor axis half-power-beam-widths) resolution.} 
\label{fig:ESO149-G003_PV-diagrams}
\end{figure*}

\begin{figure*}
 \includegraphics[width=0.99\textwidth]{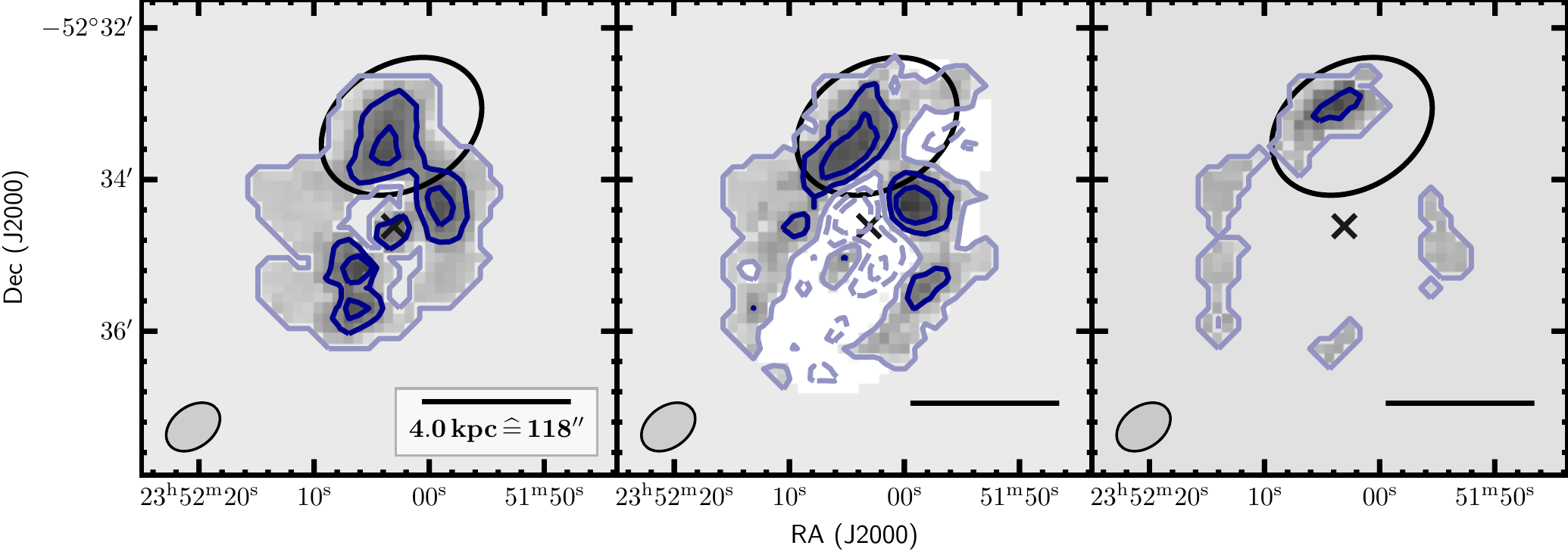}
 \caption{Difference total-intensity maps as derived from {\sc TiRiFiC} model and observed data cube. Contours denote the $-0.5,-0.25, 0., 0.25,0.5\,$-$\,M_\odot\,\mathrm{pc}^{-2}$-levels. Dashed contours show negative column-density levels, light blue corresponds to $0\,M_\odot\,\mathrm{pc}^{-2}$ Left panel: moment-0 derived from the difference data cube (observed - model). Middle panel: difference of the moment-0 of observation and model as shown in the left panel of Fig.~\ref{fig:ESO149-G003_mom0_mom1_vf}. Right panel: Moment-0 taken from the data cube applying mask as used for deriving the moment-0 shown in the left panel of Fig.~\ref{fig:ESO149-G003_mom0_mom1_vf} and additionally masking out the model data cube.}
\label{fig:ESO149-G003_mom0diffs}
\end{figure*}

In all figures in this paper, pink contours denote the model, blue contours the original observation. In addition to a comparison of model- and original datacube (Figs.~\ref{fig:ESO149-G003_data_cube_12} and \ref{fig:ESO149-G003_data_cube_24}) and original and model moment maps (Fig.~\ref{fig:ESO149-G003_mom0_mom1_vf}), we also show a set of position-velocity diagrams in Fig.~\ref{fig:ESO149-G003_PV-diagrams}, along the major and minor axes and two lines parallel to the minor axis as indicated by the arrows shown in Fig.~\ref{fig:ESO149-G003_mom0_mom1_vf} (left). We highlight that the kinematics of the galaxy is modeled by only four data points, one rotation velocity, three data points describing the dispersion profile. Given that the kinematics of the galaxy is very well resolved (see ellipses in the lower left corners in Fig.~\ref{fig:ESO149-G003_PV-diagrams} describing the resolution of our observations) the model is well constrained. The comparison shows a remarkable resemblance of the simple model to the data, in particular when focusing on only the morphological distribution of the \ion{H}{i}. As expected for a kinematically symmetrical model the velocity field is not as well represented as the total-intensity distribution (Fig.~\ref{fig:ESO149-G003_mom0_mom1_vf}). A closer inspection of the cube and the position-velocity diagrams shows that the lower-level intensity is shifted towards anomalous velocities, causing the discrepancy, while at higher level a good resemblance is achieved. Additionally, the cube and a slice perpendicular to the major axis in the North of the galaxy show clearly the characteristics of the extra-component as not being part of a regularly rotating disc (slice B in Figs.\ref{fig:ESO149-G003_mom0_mom1_vf}, left, and \ref{fig:ESO149-G003_PV-diagrams}).\\
\indent
A further outcome of the modelling process is the coincidence of the optical centre of the galaxy and the kinemato-morphological centre of the \ion{H}{i} model. The white crosses in all images denote the model centre and not the optical centre, which is separated by $5.\!\!\arcsec2$ only, well below the nominal resolution of the \ion{H}{i} observations. This is a further indication that the model optimally resembles the regular \ion{H}{i} component in the galaxy.\\
\indent
To get an idea about the location and total mass of anomalous \ion{H}{i} in galaxies is a notoriously difficult task, especially for poor spatial resolution observations where the anomalous gas appears (in projection) connected to the main body of the galaxy. Because the galaxy is highly inclined emission from gas connected to the Northern extension, which lies behind the disc, might mix in projection with emission coming from the whole North-West part of the galaxy.
To attempt bracketing the potential amount of anomalous gas and again confirm the validity of our model, we calculate the apparent mass of the anomalous gas using different methods.\\
\indent
In a first approach we subtract the model data cube from the original data cube voxel by voxel, to then calculate the moment-0 map of the residual data cube using the same convolution-and-clipping approach as described in $\S$\ref{sect:HI_observations}. The result is shown in the left panel of  Fig.~\ref{fig:ESO149-G003_mom0diffs}. While the extra-component to the North is clearly standing out, in this image it appears that also to the South and the centre of the galaxy, some significant amount of anomalous gas (as compared to our model) is present, turning up as peaks in our difference map. In addition it appears that some gas to the North-West stands out with respect to the model which might be connected to the Northern extension, but lies in front of the galaxy. The total mass of the "excess" \ion{H}{i} is $5.4\cdot 10^6\,M_\odot$, roughly 8\% of the total \ion{H}{i} mass ($7.1\cdot 10^7\,M_\odot$). In the area of the Northern extension (inside the ellipse shown in Fig.~\ref{fig:ESO149-G003_mom0diffs}) we detect $2.7\cdot 10^6\,M_\odot$.\\
\indent
The second approach consists of subtracting the total-intensity map of the model from the total-intensity map derived from the original data (using the same mask as described in $\S$\ref{sect:HI_observations}). The result is shown in the middle panel of Fig.~\ref{fig:ESO149-G003_mom0diffs}. Here, the only obvious anomalous component beneath the Northern extension is the component to the North-West, formerly connected to the Northern extension. This might suggest that the Northern extension could be part of a somewhat larger cloud complex partly covering the main body of the galaxy. The total mass as derived from the positive detected emission is $5.1\cdot 10^6\,M_\odot$, roughly 7\% of the total \ion{H}{i} mass and in rough agreement with the estimate using the first method, while in the area of the Northern extension we measure $2.4\cdot 10^6\,M_\odot$.\\
\indent
The third approach is meant to provide a confirmation of the existence of extraplanar gas in the galaxy. When calculating the total-intensity map we mask out the volume covered by the mask of the model (calculated with a pixel threshold of $0.9\,\rm{mJy}\,\rm{beam}^{-1}$, theoretically corresponding to $0.75\,\sigma_\mathrm{rms}$ in the convolved observed data cube, see $\S$\ref{sect:HI_observations}), instead of adding it. The result, shown in the right panel of Fig.~\ref{fig:ESO149-G003_mom0diffs}, contains emission solely from the Northern extension. We hence confirm the existence of the component and can estimate a lower limit to its mass,
which we calculate from this map to be $4.9\cdot 10^5\,M_\odot$ or less than 1\% of the total \ion{H}{i} mass.\\
\indent
In conclusion, while the existence of anomalous gas in ESO~149-G003 can be established by a comparison with our model, which contains only regularly rotating gas, its total mass is largely uncertain. It is, however, not a dominant fraction of the gas, which renders our model quite reliable. Assuming that our TiRiFiC model perfectly reproduces the emission only from the main \ion{H}{i} body of the galaxy, as much as 7\%--8\% of the gas in the galaxy might be external to the galaxy. It must be noted, however, that even if we model a regularly rotating disc, a best fit could still include existing anomalous gas in the galaxy, as the optimization process results in a best possible match of model and data. Our approach is most conservative in that sense. One indication that we might attribute more \ion{H}{i} than actually appropriate to the regularly rotating is the unusually high dispersion in our disc model (another is that we are not able to reproduce the details of the velocity field, merely its global gradient). Instead of a physical dispersion or random motion within the galaxy disc, the higher local dispersion might well contain a contribution from extraplanar gas which at the given resolution is indistinguishable from the main disc, or line broadening through local or global deviations from circular motion, e.g. bar streameing, which cannot be distinguished due to the limited spatial resolution. Further insight into the overall kinematics of all \ion{H}{i} components in ESO~149-G003 can only come from more sensitive observations with better angular resolution, as this would provide a higher number of independent data points and hence a more complex model that could include extraplanar gas as a separate component from the main disc.
%%%%
%%%%
%%%%%
\section{The characteristics and the origin of anomalous gas in ESO~149-G003} \label{sect:D}
In the previous section we established the presence of extraplanar and anomalous gas in ESO~149-G003. Here we briefly discuss potential interpretations of the observed kinematical deviations of the neutral gas from a common rotating disc structure, to then examine whether the galaxy follows or deviates from well-established scaling relations or shows (other) signatures which would then provide a clue about the origin of the anomalous gas.
\subsection{Morphological and kinematical structure}\label{sect:kin}
The most significant feature of the anomalous gas component is the occurrence of the Northern extension, which clearly indicates the presence of extraplanar gas in ESO 149-G003. Besides this extension, the data inspection and comparison with a model of a regularly rotating disc strongly suggests the presence of a significant amount of gas with anomalous kinematics (deviating from circular rotation) associated with ESO~149-G003 Given the spatial resolution of our MeerKAT observations, it would be premature to settle on a specific interpretation for the origin of these deviations from circularity. Such deviations are typical in dwarf galaxies especially at the low mass end \citep[e.g.][]{begum_gas_2006}, but do occur also for higher-mass dIrr galaxies \citep[e.g. CVn {\sc{i}} dwA, see Appendix D in][]{oh_high-resolution_2015}. In this respect ESO~149-G003 is not a unique exception. For the inner kinematics with the systematically tilted isovelocity contours two plausible explanations exist. If a portion of the gas or the whole of the gas moves radially with respect to the rotation axis of the galaxy a systematic shift of the radial velocities, causing the isovelocity contours to twist, could be the result. A systematic global infall or outflow is one manifestation of radial motion \citep[see e.g.][]{spekkens_modeling_2007}. A vertical infall of gas onto the mid-plane can induce a kinematic shift of the integrated spectra due to projection effects. In order to achieve such a strong kinematical twist, however, with isovelocity contours tilted by nearly $45\deg$ with respect to the morphological major axis, the velocities and the mass ratio of the gas with respect to the main body would have to be comparable to the rotation velocity of the main body. A second possibility to explain the observations are noncircular, closed orbits, i.e. streaming motions along a noncircular potential. In that case inwards and outwards motion would alternate along a circle around the kinematical axis \citep{franx_evidence_1994,schoenmakers_asymmetries_1999}. While ESO~149-G003 has been classified to be barred \citep{deVaucouleurs1991} and bar streaming motions could be responsible for a kinematic signature of this kind, the length of the kinematic twist exceeds the inner body of the galaxy, indicating a large-scale elongated potential, which might include asymmetries in the dark matter distribution of the galaxy.\\
\indent
The distinct kinematical separation of the outer component is rather peculiar and cannot be explained by an S-shaped warp, even not an extreme one, as the twist is not symmetric with respect to the centre of the galaxy. Currently we do not have a good interpretation for the kinematics in the outskirts of the galaxy. We might speculate that distinct accreting or outflowing component, of which the Northern extension is the tip of the iceberg, most distinctive with respect to the main body of the galaxy. The structure of the difference maps (see Fig.~\ref{fig:ESO149-G003_mom0diffs}) may support this picture. Whichever mechanism may be responsible for the peculiar kinematics (of the single components) of the gas in ESO-149-G003, the kinematics and the asymmetric morphology indicate that the galaxy is dynamically not settled, which is underlined by the untypically high measured gaseous dispersion (see Fig.~\ref{fig:ESO149-G003_modelpars}), unusually high even for a spiral galaxy. Such enhanced dispersion indicates either a high amount of turbulent motion in the gaseous disc or the presence of unresolved non-circular motion, further corroborated by our inability to fit the galaxy using more than one free parameter for the dispersion. A decline of the dispersion profile towards the centre is not common, but has been observed in a number of galaxies. \citet{ianjamasimanana_accurate_2017} find for a sub-sample of the the THINGS \citep{walter_things:_2008} that the \ion{H}{i} dispersion profiles of spiral galaxies are typically (exponentially) declining, while they are much flatter for dwarf galaxies. A closer look at their sample shows that an initial increase of the dispersion with radius can be found for the dwarf galaxies M81 dwA, while it would be compatible with the profiles of DDO 53 and Ho {\sc{i}}. Also the detailed analysis of the LITTLE THINGS \citep{hunter_little_2012} sample by \citet{oh_high-resolution_2015} shows dIrr galaxies with this property (IC~1613, NGC~3738, Haro~36).
\subsection{Tully-Fisher relations}\label{sect:TFR}
The Tully-Fisher relation (TFR) is a correlation of the global rotation velocity (represented by one number) with the luminosity, and is the most frequently applied scaling relation for galaxies \citep{Tully1977}. The underlying driver seems to be the correlation of the total mass with the rotation speed of the galaxy, the so-called baryonic Tully-Fisher relation \citep[e.g.][]{mcgaugh_baryonic_2005,mcgaugh_baryon_2010,mcgaugh_baryonic_2012}. We make use of it to gain an insight whether the global kinematics of ESO~149-G003 is unusual. This is possible, because the adopted distance is not a Tully-Fisher-based distance. Instead we used a distance determined using the tip of the red giant branch (TRGB) method \citep{Tully2013}. \\
\indent
For our study we selected relations from two more recent works; one by \citet{karachentsev_local_2017}, which investigates the TFRs of local dwarf galaxies and employing the integrated line width ($W_{50}$), the other by \citet{ponomareva_multiwavelength_2017,ponomareva_light_2018}, a more recent study which employs the rotation velocity of the galaxies as correlated with luminosities and mass estimates in the FIR, leading to a lower scatter. The BTFRs of both groups make use of a baryonic mass estimate of the form:
\begin{equation}
    M_\mathrm{C, \ion{H}{i}}\,=\,\gamma_\mathrm{c}L_\mathrm{c}+\eta\,M_\ion{H}{i}
\end{equation}
$\gamma_\mathrm{c}$ is a colour-dependent scaling factor for the luminosity measured in a certain band, and $\eta$ a scaling factor to account for the scaling of the \ion{H}{i} mass $M_\ion{H}{i}$ , other gas phases (in particular molecular gas), and self-absorption. To account for helium abundance, usually a factor between 1.3 and 1.4 is assumed. In our case, we neglect the mass contribution of the molecular component, found to be negligible with respect to the \ion{H}{i} mass for late-type dwarf galaxies \citep{young_ratio_1989,mcgaugh_gas_1997}.
We first turn to the K-band and baryonic Tully-Fisher relation employing the line width by \citet{karachentsev_local_2017}:
\begin{equation}
\begin{aligned}
    &\begin{aligned}
        \log\left(\frac{L^\mathrm{TFR}_\mathrm{K}}{L_\odot}\right)\,&=\,2.91\,\log\left(\frac{W^\mathrm{c}_{50}}{\mathrm{km}\,\mathrm{s}^{-1}}\right)+2.91\pm0.47\\
        &=\,7.5\,\pm\,0.5
    \end{aligned}\\
    &\begin{aligned}
        \log\left(\frac{\,M^\mathrm{TFR}_\mathrm{K, \ion{H}{i}}}{M_\odot}\right)\,&=\,2.61\,\log\left(\frac{W^\mathrm{c}_{50}}{\mathrm{km}\,\mathrm{s}^{-1}}\right)+3.64\pm 0.38\\
        &=\,7.7\,\pm\,0.4
    \end{aligned}&\\
\end{aligned}
\end{equation}
Here, $W^\mathrm{c}_{50}$ is the line width at 50\% of the peak flux density (see Table~\ref{tab:gal_prop}), corrected for inclination (i.e. divided by $\sin\,i$, where $i$ is the inclination). With this, especially in the case for dwarf galaxies, for which the internal dispersion plays a significant role, the corrected line width overestimates the edge-on line width if the dispersion is isotropic. On the other hand, an uncorrected line width would not take the galaxy rotation properly into account and underestimates the edge-on line width. As our galaxy lies close to edge-on and the correction is small, we employed the TFR involving the corrected line width, although \citet{karachentsev_local_2017} also provide a TFR without inclination correction. The observed K-band luminosity of $\log\left(\frac{L^{obs}_\mathrm{K}}{L_\odot}\right)\,=\,8.0$ taken from \citet{karachentsev_local_2017}, based on 2MASS observations, and the Baryonic mass $\log\left(\frac{\,M^\mathrm{obs}_\mathrm{K, \ion{H}{i}}}{M_\odot}\right)\,=8.2\,\pm\,0.2$, using  $\gamma_\mathrm{K}\,=\,0.6$ and $\eta\,=\,1.33$ \citep{karachentsev_local_2017} and our own measured \ion{H}{i} mass, fall below the predicted quantities by roughly $1\,\sigma$, where $\sigma$ is the scatter (including a slight contribution from the error of our observables). The pure stellar mass, using $\gamma_\mathrm{K}\,=\,0.6$ is $M^\mathrm{obs}_\mathrm{K}\,=\,5.4\cdot10^7\,M_\odot$.\\
\indent
Employing the TFRs by \citet{ponomareva_multiwavelength_2017,ponomareva_light_2018} and using the asymptotic velocity $v_\mathrm{flat}\,=\, v_\mathrm{rot}(r_\ion{H}{i})\,=\,18.3\,\pm\,1.4$ (see Table~\ref{tab:gal_prop} and Fig.~\ref{fig:ESO149-G003_modelpars}), we calculate the expected $3.6\,\mathrm{\mu m}$ luminosity and the baryonic mass as:
\begin{equation}
    \begin{aligned}
    &\begin{aligned}
        \log\,\left(\frac{L^\mathrm{TFR}_{3.6\mathrm{\mu}}}{L_\odot}\right)\,&=\,(3.7\pm0.11)\,\log\,\left(\frac{2v_\mathrm{flat}}{\mathrm{km}\,\mathrm{s}^{-1}}\right)+1.3\pm 0.3\\
        &=\,7.1\,\pm\,0.4\\
    \end{aligned}\\
     &\begin{aligned}
       \log\,\left(\frac{M^\mathrm{TFR}_{3.6\mathrm{\mu, \ion{H}{i}}}}{M_\odot}\right)\,&=\,(2.99\pm0.22)\,\log\,\left(\frac{2v_\mathrm{flat}}{\mathrm{km}\,\mathrm{s}^{-1}}\right)+2.88\pm 0.56\\
        &=\,7.6\,\pm\,0.7
    \end{aligned}
    \end{aligned}
\end{equation}\label{eq:spitmas}
To compare to observed luminosities, we first attempt to use the 
$\mathrm{3.6\mu}$ 
absolute magnitude of 
$M_{3.6\mathrm{\mu}}\,=\,-13.\!\!^\mathrm{m}273$ 
from the Spitzer Survey of Stellar Structure in Galaxies (S$^4$G) catalogue \citep[in the NASA/IPAC Infrared Science Archive {\sc IRSA}\footnote{\url{https://irsa.ipac.caltech.edu}}]{sheth_spitzer_2010,munoz-mateos_spitzer_2015}. Following \citet{ponomareva_multiwavelength_2017}, we use  $M_\odot(3.6\mathrm{\mu})\,=\,3.\!\!^\mathrm{m}24$
\citep{oh_high-resolution_2008,ponomareva_multiwavelength_2017} to derive the luminosity and then scale the luminosity from the distance of 
$6.25\,\mathrm{Mpc}$
\citep{munoz-mateos_spitzer_2015} to the distance of $7.0\,\mathrm{Mpc}$. We obtain 
$\log\left(\frac{L^\mathrm{obs}_{3.6\mathrm{\mu}}}{L_\odot}\right) =\,6.7$
or 
$L^\mathrm{obs}_{3.6\mathrm{\mu}}\,=\,5.1\cdot10^6\,L_\odot$. Using 
$\gamma_\mathrm{3.6\mathrm{\mu}}\,=\,0.35$
and 
$\eta\,=\,1.4$
\citep{ponomareva_multiwavelength_2017} we calculate $\log\left(\frac{M^\mathrm{obs}_{3.6\mathrm{\mu}}}{M_\odot}\right)\,=8.0\,\pm\,0.2$
and $M^\mathrm{obs}_{3.6\mathrm{\mu}}\,=\,1.8\cdot10^6\,M_\odot$. 
The derived stellar mass stands in contrast to the K-band stellar mass $M^\mathrm{obs}_\mathrm{K}\,=\,5.4\cdot10^7\,M_\odot$ using the prescription of \citet[see above]{karachentsev_local_2017}. We hence re-visit the S$^4$G catalogue \citep{sheth_spitzer_2010,munoz-mateos_spitzer_2015} to use the listed infrared stellar mass of  $\log\left(\frac{M^\mathrm{obs}_{\mathrm{S}^4\mathrm{G, ori}}}{M_\odot}\right)\,=\,7.575$,
and scale from a distance of $6.25\,\mathrm{Mpc}$ \citep{munoz-mateos_spitzer_2015} to the distance of 
$7.0\,\mathrm{Mpc}$. 
We obtain 
$\log\left(\frac{M^\mathrm{obs}_{\mathrm{S}^4\mathrm{G}}}{M_\odot}\right)\,=\,7.675$
or 
$M^\mathrm{obs}_{\mathrm{S}^4\mathrm{G}} \,=\, 4.7\cdot 10^7 M_\odot$,
in very good agreement with $M^\mathrm{obs}_\mathrm{K}\,=\,5.4\cdot10^7\,M_\odot$ (or, alternatively, still with $M^\mathrm{obs}_\mathrm{FIR}\,=\,3.6\cdot10^7\,M_\odot$ as derived by \citet{wang_local_2017} using WISE \citep{wright_wide_field_2010} data). The origin of the discrepancies between the $3.6\mathrm{\mu}$ luminosities and stellar masses between \citet{ponomareva_light_2018} and \citet{munoz-mateos_spitzer_2015} is not clear. We, however, assume it to be a safe approach to adopt the catalogued stellar mass from \citet{munoz-mateos_spitzer_2015} instead of deriving it ourselves using a prescription.\\
). Accordingly, we derive for the total observed mass 
$\log\left(\frac{M^\mathrm{obs}_{\mathrm{S}^4\mathrm{G},\ion{H}{i}}}{M_\odot}\right)\,=\,8.2\,\pm\,0.2$.\\
\indent
All TFR luminosities and baryonic masses are hence in reasonable agreement with observed luminosities (and masses). We can interpret the fact that using the line width results in a slight overprediction of both luminosity and stellar mass as an independent confirmation of the increased \ion{H}{i} dispersion in the galaxy. The galaxy does, however, not fall off a scaling relation connecting the overall kinematics and luminosity, and baryonic mass content of ESO~149-G003 do not indicate that the galaxy is dynamically strongly disturbed (again justifying the use of a tilted-ring model to describe the kinematics of the galaxy).
%%%
%%%
\subsection{Neutral-hydrogen mass-size relationship}\label{sect:msr}
First observed by \citet{broeils_short_1997} the correlation between the mass and the size of the \ion{H}{i} component in galaxies has since been established as one of the tightest fundamental scaling laws \citep{ verheijen_ursa_2001, begum_figgs:_2008, swaters_westerbork_2002, noordermeer_westerbork_2005, wang_bluedisks_2013, wang_new_2016}. Thanks to our tilted-ring modelling we can determine the \ion{H}{i} diameter of $5.9\,\pm\,0.1\,\mathrm{kpc}$ from the observations. We compare this to the expected \ion{H}{i} radius following \citet{wang_new_2016} and find
\begin{equation}
\begin{aligned}
&\begin{aligned}
\log\left(\frac{D^\mathrm{pred}_\ion{H}{i}}{\mathrm{pc}}\right) \,&=\, 0.506 \log \left(\frac{M_\ion{H}{i}}{M_\odot}\right)-3.293\,\pm\,0.06\\
&=\,0.679\,\pm\,0.06
\end{aligned}\\
&\begin{aligned}
D^\mathrm{pred}_\ion{H}{i}\,&=\, %4.78\,\pm\,0.66\,\mathrm{kpc}\qquad\mathrm{,}
4.8\,\pm\,0.7\,\mathrm{kpc}\qquad\mathrm{,}
\end{aligned}
\end{aligned}
\end{equation}
whereas $\log\left(\frac{D^\mathrm{obs}_\ion{H}{i}}{\mathrm{pc}}\right) \,=\,0.771\,\pm\,0.008$. The observed size of the \ion{H}{i} disc is $~1.5\,\sigma$ on the high side as compared to the expected size of the disc, based on the \ion{H}{i} mass of the galaxy. While this might indicate that a portion of the \ion{H}{i} in the galaxy is on slightly larger orbits as compared to an average galaxy with the same \ion{H}{i} mass (has a higher specific angular momentum), and hence an extragalactic component, the evidence for this is not compelling. We cannot confirm the observation by \citet{wang_new_2016} that the galaxy falls off the \ion{H}{i} mass-size relation, which they already attributed to the poor resolution in their observations.
\subsection{Star formation and neutral hydrogen content}\label{sect:sfe}
The observed disturbances to the overall kinematics and the presence of extraplanar neutral gas in ESO~149-G003 indicates either an early stage of an interaction or a recent interaction of the galaxy with its environment. A clue about which of the scenarios is at work can be sought looking at the star formation properties. An enhanced star formation rate or even a starburst could be an indicator of a past interaction, in which either tidal forces or stellar feedback removed neutral gas from the ISM into the IGM. Conversely one would interpret the absence of any signature of such enhancement as a potential signature of an early stage of an interaction, where the extraplanar gas has not yet been accreted onto the disc, or a very weak interaction. With the star formation rates of 
$\log\left( \frac{SFR_\mathrm{H\alpha}}{M_\odot\,\mathrm{yr}^{-1}}\right) = -2.23$ as derived by \citet{karachentsev_star_2013} using 
$\mathrm{H\alpha}$ observations taken with the SAO 6m telescope \citep{kaisin_h_2007}
and
$\log\left( \frac{SFR_\mathrm{UV,FIR}}{M_\odot\,\mathrm{yr}^{-1}}\right) = -1.99$ as derived by \citet{wang_local_2017} using {\it GALEX} FUV\citep{martin_galaxy_2005,gil_de_paz_galex_2007,lee_galex_2011} {\it WISE} FIR \citep{wright_wide_field_2010} data, ESO~149-G003 does not exhibit any significantly enhanced star formation \citep[see e.g.][]{hunter_star_2004}.
and
However, the neutral gas mass in the galaxy is high. 
While a ratio of the \ion{H}{i} mass to the blue stellar luminosity of
$M_\ion{H}{i}\,L^{-1}_\mathrm{B}\,=\, 0.9\,M_\odot\,L^{-1}_\odot$ 
is not unusual, the ratio of total gas mass (by scaling $M_\ion{H}{i}$ with a factor of $1.33$ to total mass with $\log\left(M_\mathrm{gas}\,M^{-1}_\mathrm{bar}\right)\,=\,-0.2$
(using $M^\mathrm{obs}_{\mathrm{S}^4\mathrm{G, \ion{H}{i}}}$  puts it on the higher gas mass edge \citep[see e.g.][]{hunter_star_2004}. ESO~149-G003 is gas-dominated, the ratio of the gas mass to the stellar mass is $M_\mathrm{gas}\,M^{-1}_\mathrm{\mathrm{S}^4\mathrm{G}}\,=1.7\,$). It consumes its gas, however, at a normal rate, as the star formation efficiencies of $\log\left(\frac{SFE_\mathrm{H\alpha}}{\mathrm{yr}^{-1}}\right)\,=\,\log\left(\frac{SFR_\mathrm{H\alpha}\,M^{-1}_\ion{H}{i}}{\mathrm{yr}^{-1}}\right)\,=\,-10.1$ and $\log\left(\frac{SFE_\mathrm{UV,FIR}}{\mathrm{yr}^{-1}}\right)\,=\,\log\left(\frac{SFR_\mathrm{UV,FIR}\,M^{-1}_\ion{H}{i}}{\mathrm{yr}^{-1}}\right)\,=\,-9.8$ show. \citet{wong_characterizing_2016} derived $\log\left(\frac{SFE_\mathrm{UV}}{\mathrm{yr}^{-1}}\right)\,=\,-9.7\pm0.3$ studying the SINGG and SUNGG surveys \citep{meurer_survey_2006,wong_star_2007} as an average value for any type of galaxy, very close to our result.\\
\indent
Concerning resolved star formation properties, ESO~149-G003 does not seem to be extraordinary either. \citet{ryan-weber_intergalactic_2004} observed the H$\alpha$ emission in its stellar disc to be filamentary. Remarkably, ESO~149-G003 might be associated to an intergalactic \ion{H}{ii} region, separated by $1.\!\!^\prime5$ to the West \citep{ryan-weber_intergalactic_2004,torres-flores_star_2009,koribalski_local_2018}. As this region, however, has a significantly different recession velocity ($949\,\mathrm{km}\,\mathrm{s}^{-1}$), its connection to ESO149-G003 is unclear.
Given the star formation properties of ESO~149-G003 we deem it unlikely that the anomalous gas originates from stellar feedback or that we are seeing gaseous outflows from star formation regions. However, detailed, deep photometry and spectroscopy, studying the full star formation history of the object, is required to draw final conclusions in this respect.
\subsection{The origin of the anomalous gas in ESO~149-G003}\label{sect:originggas}
\indent
ESO~149-G003 seems to be rather isolated. According to \citet{koribalski_local_2018}, the galaxy lies at the edge of the Sculptor group, while \citet{karachentsev_distances_2003} classify it as a background object to the Sculptor group \citep[see also][]{ryan-weber_intergalactic_2004}. This is corroborated by the fact that the independent TRGB distance measurement by \citet{Tully2013} puts it at a greater distance than assumed by \citet[see their Fig.~5]{karachentsev_distances_2003}. \citet{nicholls_small_2011} claim that ESO~149-G003 has likely undergone no interaction with any known galaxy in its surroundings within a Hubble time.\\
\indent
Hence, should the anomalous gas in ESO~149-G003 originate from an interaction, it is most likely that it has been a past (minor) merger, not a tidal interaction. Otherwise the other studies would not indicate such a high degree of isolation. 
Based on r-band photometry by \citet{ryan-weber_intergalactic_2004}, \citet{ryan-weber_intergalactic_2004} and \citet{koribalski_local_2018} conclude that ESO~149-G003 shows optical signatures of a past interaction.
If so, that merger is, however, in a quite advanced stage, as we do not find any isolated remnant of the interaction partner. Only the potential existence of some shells in the stellar body is reported and visible. We hence conclude that if the anomalous gas in ESO~149-G003 stems from an interaction, that interaction has likely been a past merger, which would now be in its final stages. Given the advanced merger state, assuming that the gas originated from that merger, we conclude it would most likely returning, not still leaving the galaxy. The required assumptions are that the two progenitor studies indicating isolation are correct and that, with the available photometry, we would be able to identify a nearby interaction partner large enough to induce the observed kinematic anomalities and that gas that close to a galaxy would rather return than leave in an advanced merger.\\
\indent
As elaborated in the previous section, with the current star formation rate, an outflow origin of the anomalous gas is not probable.
We hence deem it more likely that ESO~149-G003 is currently accreting gas rather than expelling neutral gas.
The accreted gas is either returning gas of a past merger event or pristine gas from the IGM.
Utilizing several scaling relations we have established that ESO~149-G003 is not likely undergoing a dramatic merging or star forming event. It hence appears that the anomalous gas around the galaxy is not accreted at a sufficient rate to cause significantly enhanced star formation at the current time, although at a later stage the accretion might evolve into a starburst.
The amount of accreting gas cannot be determined as higher resolution observations would be necessary to distinguish between global radial and vertical in- or outflow in the galaxy or the ellipticity of orbits in the galaxy, but we interpret the existence of
the Northern extension of the \ion{H}{i} body, with a mass of at least 
$4.9\cdot 10^5\,M_\odot$ (less than 1\% of the total \ion{H}{i} mass),
as evidence that extraplanar gas is present in ESO~149-G003. Estimating the total mass of gas with anomalous velocities, as identified in contrast to our model, we derive a lower limit of $5.1-5.4\cdot 10^6\,M_\odot$, roughly 7\%--8\% of the total \ion{H}{i} mass ($7.1\cdot 10^7\,M_\odot$). As the literature lacks attempts to quantify the mass of anomalous components in galaxies similar to ours, it is hard to estimate whether ESO~149-G003 is peculiar in this respect. \citet{marasco_halogas_2019} derive a typical anomalous gas fraction of 10\%--15\% for 15 HALOGAS \citep{heald_westerbork_2011} galaxies, which would render ESO~149-G003 a rather average galaxy, while a similar study is missing for dwarf galaxies. \\
Concerning the nature of the accreting gas, with the observations at hand, we may speculate. Comparable cases for potential interactions of isolated dIrr galaxies with the IGM are rare. \citep{de_blok_meerkat_2020} present MeerKAT observations of the dwarf galaxy ESO~302-G014 and show that the isolated galaxy with an \ion{H}{i} mass of $3.8\cdot10^8\,\,M_\odot$ contains a filamentary extension of $4.6\cdot10^6\,\,M_\odot$, potentially not unlike the one observed for ESO~149-G003, and constituting a similar fraction of \ion{H}{i} as compared to the total \ion{H}{i} mass. While the authors do not exclude pristine accretion, they favour the interaction with a small dwarf to be responsible for the occurrence of the filament. While ESO~149-G003 might be a similar case, one would need to scale the mass of such a gas donor.
As already stated above, based on r-band photometry by \citet{ryan-weber_intergalactic_2004}, \citet{ryan-weber_intergalactic_2004} and \citet{koribalski_local_2018} conclude that ESO~149-G003 shows signs of a past interaction in the optical. \citet{koribalski_local_2018} hence suggest that the extraplanar gas in ESO~149-G003 stems from an accretion event, it has to be accretion of a significantly smaller object than the dwarf galaxy itself. While objects of such small mass do exist \citep[e.g.][]{ryan-weber_local_2008}, in this scenario it would be surprising that ESO~149-G003 appears not to show any signs of recent enhanced star formation. Furthermore, it is unclear how frequent mergers of isolated dwarf galaxies take place. In fact, isolated dwarf galaxies can show rather peculiar kinematics \citep[][]{kreckel_kk_2011,kreckel_void_2016}, while to date it is not clear whether these are due to past mergers or the accretion of pristine gas. Also, associations of very low-mass gas-rich galaxies have been observed \citep{ball_enigmatic_2018}, but they are very rare. A pristine accretion event can hence not be excluded, making ESO~149-G003 an object of high interest.
Concerning the fate of ESO 149-G003, it may be assumed that the accretion of the gas cloud leads to an enhanced star formation. It is, in fact, slightly elevated as compared to the average dIrr \citep{van_zee_evolutionary_2001}. In a sample of 18 BCDs, \citet{lelli_triggering_2014} identified three systems, for which they do not exclude pristine gas accretion to be a potential trigger for a starburst phase. In fact the enhanced measured dispersion, potentially connected to bar-like motion in the galaxy might lead to an inflow of gas triggering star formation at a later stage. A similarly high dispersion, in addition to filamentary structures in the galaxy outskirts has been measured for the post-starburst galaxy NGC~1569 \citep{johnson_stellar_2012,oh_high-resolution_2015}. To be more conclusive, follow-up observations at various wavelengths, but in particular in the \ion{H}{i} component will be required.
\section{Summary and outlook} \label{sect:SaO}
In this paper we presented MeerKAT \ion{H}{i} observations of the dwarf galaxy ESO~149-G003. We summarize our findings as follows:
\begin{itemize}
    \item The sensitivity of the observations allows us to identify kinematically anomalous gas in the galaxy.
    \item Using a tilted-ring model, we are able to reproduce the column-density distribution in the galaxy, but we cannot reproduce its complete kinematics.
    \item We confirm the existence of an extraplanar component in the galaxy to the North.
    \item We estimate 7\%--8\% as a rough lower limit for the fraction of anomalous gas in the galaxy.
    \item While ESO~149-G003 is gas-dominated, its kinematical and photometrical properties are consistent with various scaling relations.
    \item Given the isolation of the galaxy we conclude that ESO~149-G003 is likely accreting neutral gas, but neither the accretion rate nor the origin of the gas can be determined with any precision. Both a merger as well as the accretion of pristine gas are possible as the origin of the extraplanar gas.
\end{itemize}
 Given its dynamical state, ESO~149-G003 is hence a very interesting object, as we might be observing the accretion (or re-accretion) of neutral gas onto a dwarf irregular galaxy. How many objects like this will be observed with the imminent surveys with SKA precursors remains to be seen. While such observations will provide us with good statistics, case studies of such presumably rare events are still useful to understand the precise mechanisms behind gas accretion in dwarf galaxies.
 Extrapolating from this study, based on observations conducted with 16 out of 64 antennas, MeerKAT-64, especially in spectral zoom mode, will provide the sensitivity and the spectral resolution to be an ideal instrument for the study of the detailed gas kinematics in dwarf galaxies.
 Deeper \ion{H}{i} observations required to further model the galaxy kinematics as well as deeper photometry is hence not very hard to achieve in the near future. We hence suggest that ESO~149-G003 is a very good candidate to study anomalous gas in dwarf galaxies.
 %%%
\section*{Acknowledgements}
This paper is dedicated to all the workers who built the MeerKAT telescope and made this study possible.\\
\indent We thank Fernando Camilo (SARAO) and Dominik J. Bomans (RUB) for very helpful comments and discussions.\\
\indent
The MeerKAT telescope is operated by the South African Radio Astronomy Observatory, which is a facility of the National Research Foundation, an agency of the Department of Science and Innovation.\\
\indent This work is partly based on data obtained with the MeerLICHT telescope, located at the SAAO Sutherland station, South Africa. The MeerLICHT telescope is run by the MeerLICHT consortium, on behalf of Radboud University, the University of Cape Town, the Netherlands Foundation for Scientific Research (NWO), the National Research Facility of South Africa through the South African
Astronomical Observatory, the University of Oxford, the University of Manchester and the University of Amsterdam.\\
\indent Part of the data published here have been reduced using the CARACal pipeline, partially supported by ERC Starting grant number 679629 “FORNAX”, MAECI Grant Number ZA18GR02, DST-NRF Grant Number 113121 as part of the ISARP Joint Research Scheme, and BMBF project 05A17PC2 for D-MeerKAT. Information about CARACal can be obtained online under the URL:: \url{https://caracal.readthedocs.io}.\\
\indent PK and RJD acknowledge support from the BMBF project  05A17PC2 for D-MeerKAT.\\
This project has received funding from the European Research Council (ERC) under the European Union's Horizon 2020 research and innovation programme (grant agreement no. 679629; project name FORNAX)\\
\indent PAW acknowledges research funding from NRF and UCT.\\
\indent KP acknowledges funding by the National Astrophysics and Space Science Programme (NASSP), the National Research Foundation of South Africa (NRF) through a South African Radio Astronomy Observatory (SARAO) bursary, and the University of Cape Town (UCT) for work on MeerLICHT.\\
\noindent We acknowledge the use of computing facilities of the Inter-University Institute for Data Intensive Astronomy (IDIA) for part of this work (the reduction of the MeerLICHT data). IDIA is a partnership of the Universities of Cape Town, of the Western Cape and of Pretoria.\\
\indent This research has made use of the NASA/ IPAC Infrared Science Archive, which is operated by the Jet Propulsion Laboratory, California Institute of Technology, under contract with the National Aeronautics and Space Administration.\\
\indent
The Digitized Sky Surveys were produced at the Space Telescope Science Institute under U.S. Government grant NAG W-2166. The images of these surveys are based on photographic data obtained using the Oschin Schmidt Telescope on Palomar Mountain and the UK Schmidt Telescope. The plates were processed into the present compressed digital form with the permission of these institutions.\\
\indent This work has made use of data from the European Space Agency (ESA) mission {\it Gaia} (\url{https://www.cosmos.esa.int/gaia}), processed by the {\it Gaia}
Data Processing and Analysis Consortium (DPAC,
\url{https://www.cosmos.esa.int/web/gaia/dpac/consortium}). Funding for the DPAC
has been provided by national institutions, in particular the institutions
participating in the {\it Gaia} Multilateral Agreement.
%
%%%%%%%%%%%%%%%%%%%%%%%%%%%%%%%%%%%%%%%%%%%%%%%%%%
%
\section*{Data availability}
The data underlying this article will be shared on reasonable request to the corresponding author.
\bibliographystyle{mnras}
\bibliography{hi_in_eso149_g003} % if your bibtex file is called example.bib

%%%%%%%%%%%%%%%%%%%%%%%%%%%%%%%%%%%%%%%%%%%%%%%%%%

%%%%%%%%%%%%%%%%% APPENDICES %%%%%%%%%%%%%%%%%%%%%

\appendix
\section{Images and models}
We provide additional images for the perusal of the interested reader in this appendix.
\begin{figure*}
 \includegraphics[width=0.85\textwidth]{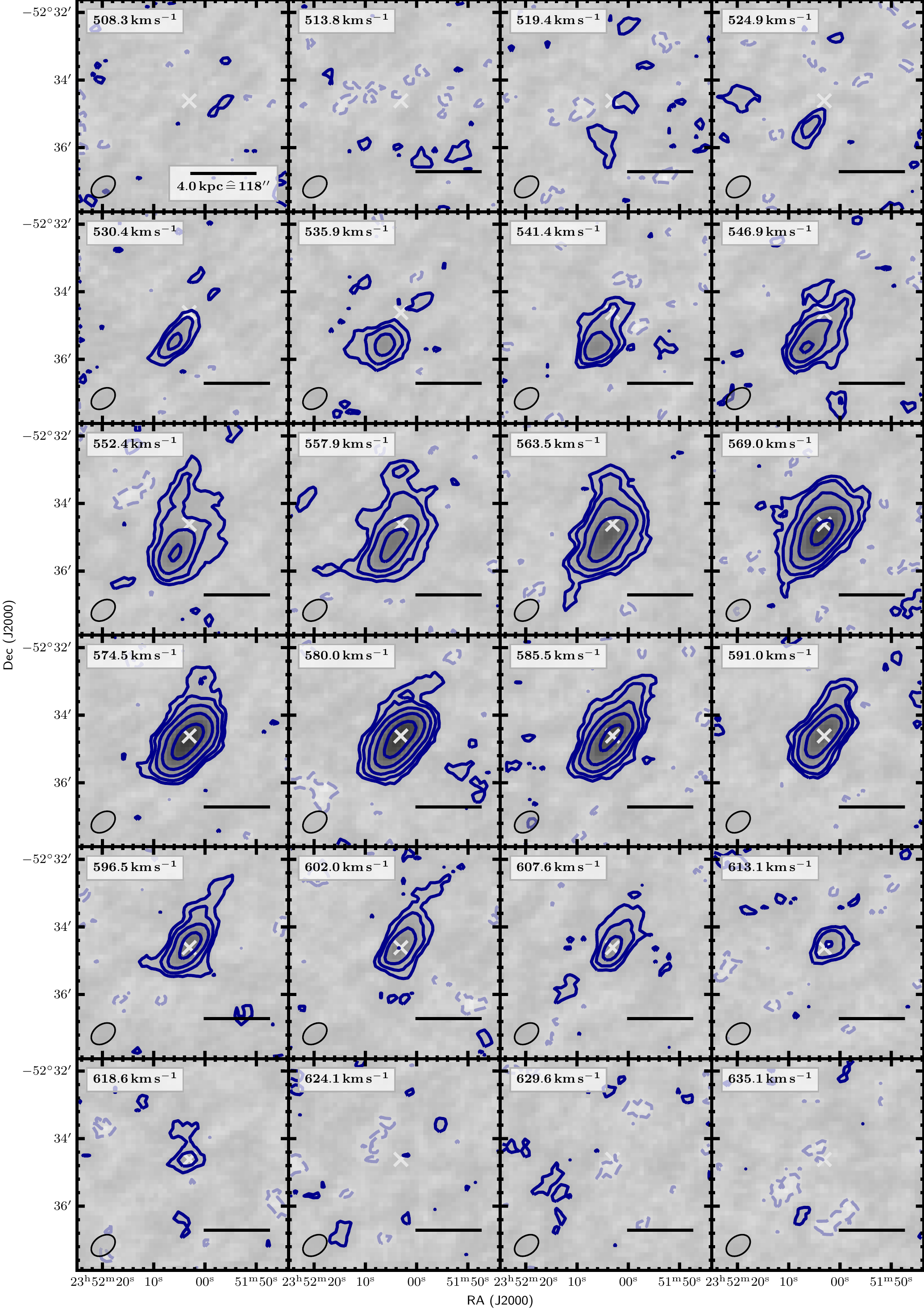}
 \caption{Observed MeerKAT \ion{H}{i} data cube. Contours denote the $-2,2,4,8,16,32\,-\,\sigma_\mathrm{rms}$-levels, where $\sigma_\mathrm{rms}\,=\,1\,\mathrm{mJy}\,\mathrm{beam}^{-1}$. The white cross represents the kinematical centre.}
 \label{fig:ESO149-G003_data_cube_24_obs}
\end{figure*}
%%%%
\begin{figure*}
 \includegraphics[width=0.85\textwidth]{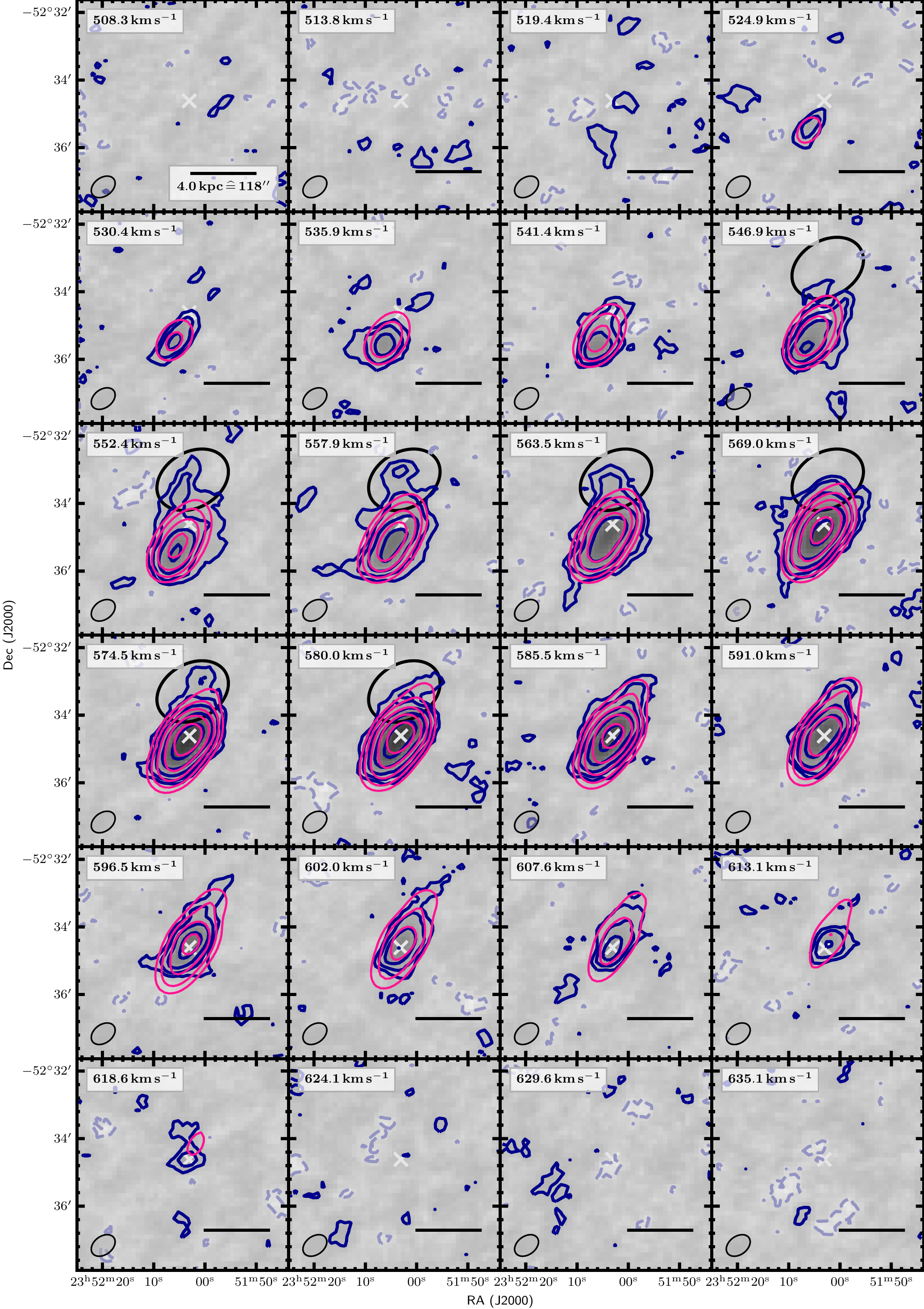}
 \caption{MeerKAT \ion{H}{i} data cube. Contours denote the $-2,2,4,8,16,32\,-\,\sigma_\mathrm{rms}$-levels, where $\sigma_\mathrm{rms}\,=\,1\,\mathrm{mJy}\,\mathrm{beam}^{-1}$ . Blue: the observed data cube. Pink: the {\sc TiRiFiC} model. Dashed lines represent negative intensities. The white cross represents the kinematical centre of the model, the ellipse to the lower left the syntesized beam ($HPWB$), the black ellipse to the top highlights kinematically anomalous gas.}
 \label{fig:ESO149-G003_data_cube_24}
\end{figure*}
%%%%
\begin{figure*}
 \includegraphics[width=\textwidth]{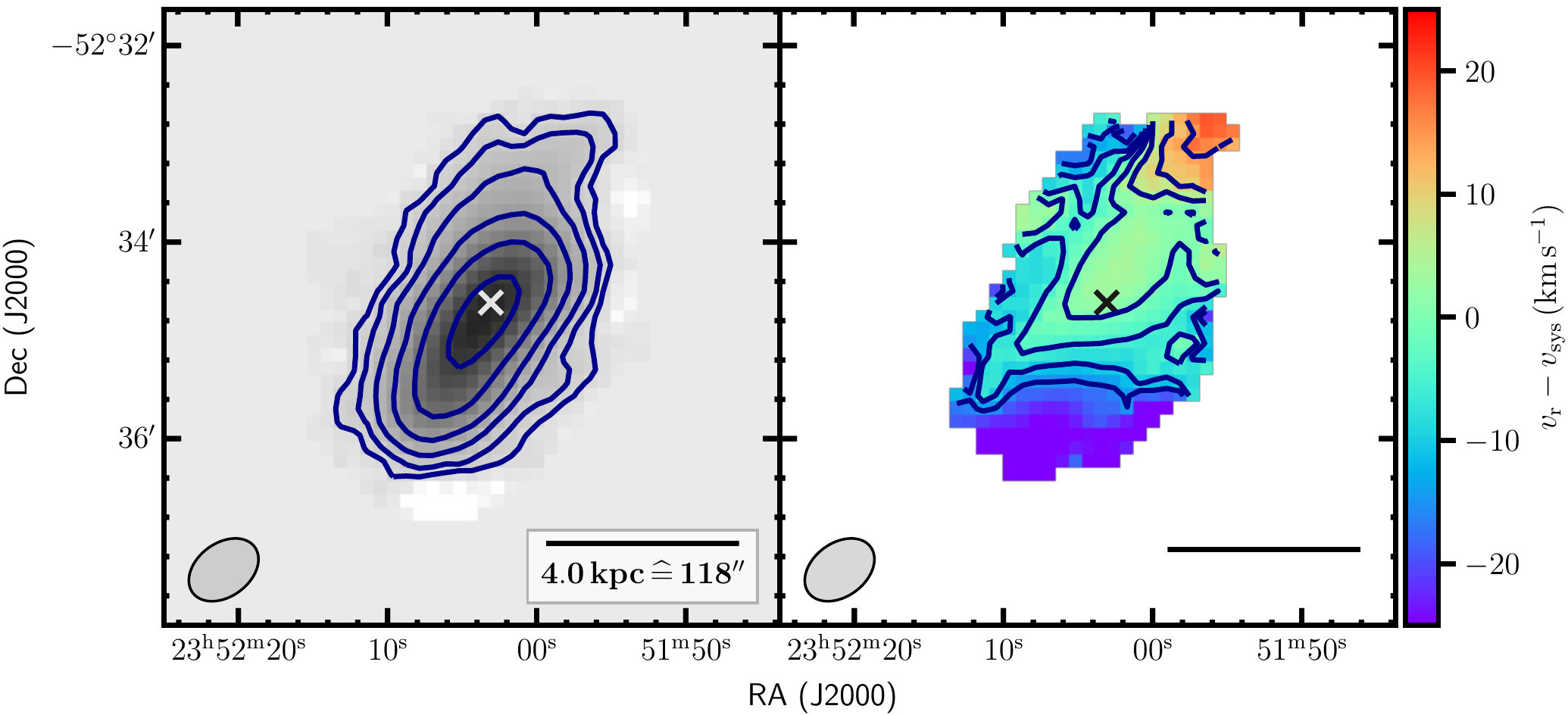}
 \caption{MeerKAT \ion{H}{i} total-intensity map and moment-1 velocity field, as derived from the observed data cube shown in Figs.~\ref{fig:ESO149-G003_data_cube_12}, \ref{fig:ESO149-G003_data_cube_24}, and \ref{fig:ESO149-G003_data_cube_24_obs}. The crosses represent the kinematical centre, the ellipse to the lower left the syntesized beam ($HPWB$). Left: Total-intensity map. Contours denote the $0.25,0.5,1,2,4,8\,$-$\,M_\odot\,\mathrm{pc}^{-2}$-levels. Right: velocity field. Contours are isovelocity contours. They denote the $-15, -10, -5, 0, 5, 10,15\,\mathrm{km}\,\mathrm{s}^{-1}$-levels relative to the systemic velocity $v_\mathrm{sys}\,=\,578\,\mathrm{km}\,\mathrm{s}^{-1}$.}
\label{fig:ESO149-G003_mom0_mom1_vf_obs}
\end{figure*}
%%%%
\begin{figure*}
\begin{center}
 \includegraphics[width=\columnwidth]{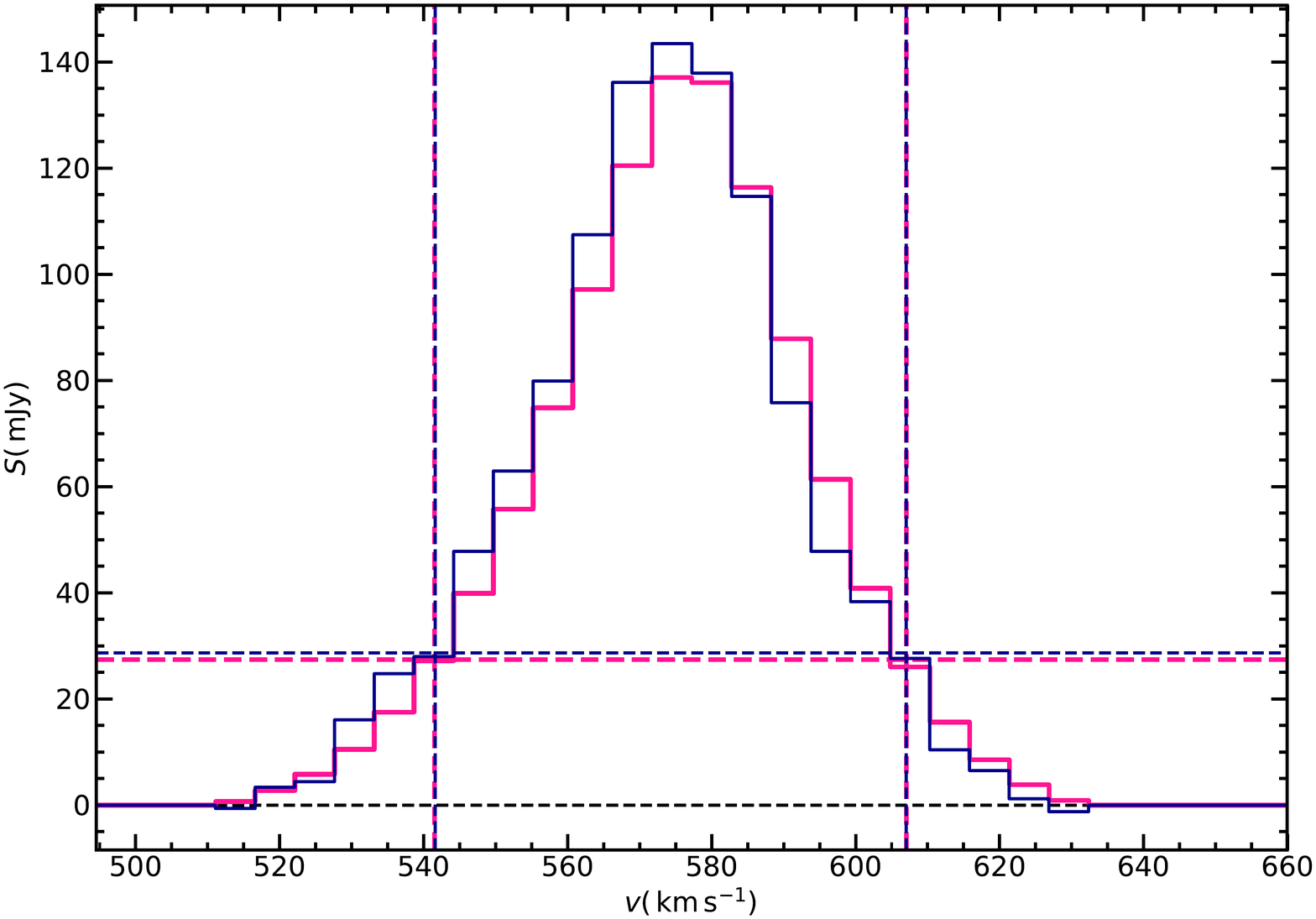}
 \end{center}
 \caption{\ion{H}{i}-spectrum of ESO~149-G003. Blue: observed. Pink: {\sc TiRiFiC} model. The horizontal lines denote 20\% of the peak, the vertical lines the corresponding velocities.}
\label{fig:ESO149-G003_spectrum}
\end{figure*}
%%%%
\begin{figure*}
\begin{center}
 \includegraphics[width=\columnwidth]{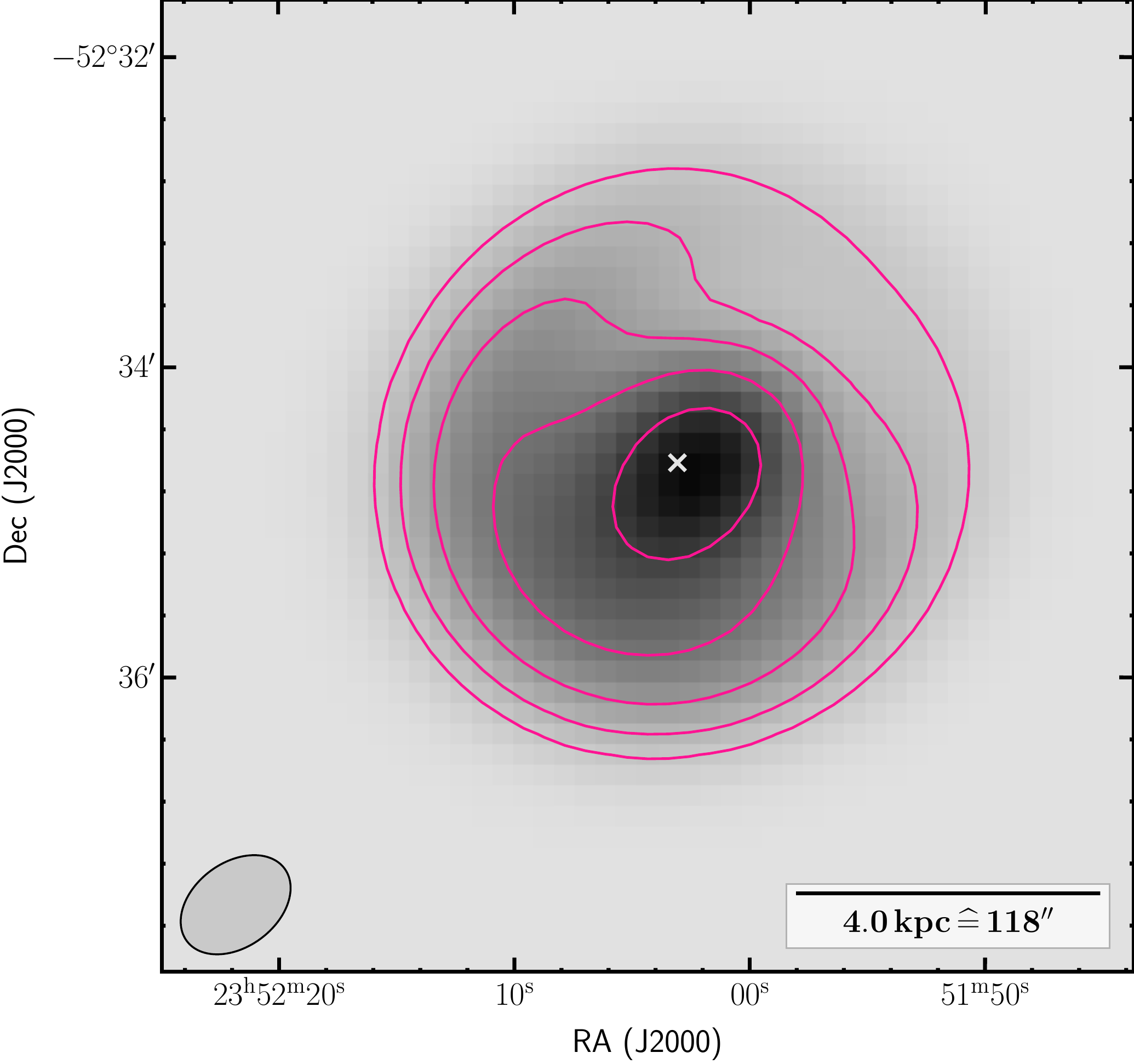}
 \end{center}
 \caption{Face-on total-intensity map of the {\sc TiRiFiC} model. The \ion{H}{i} disc is simulated to be seen face-on, but at the same distance and the same resolution as compared to the observations. Contours: $0.25,0.5,1,2,4\,$-$\,M_\odot\,\mathrm{pc}^{-2}$.}
\label{fig:ESO149-G003_topview_mom0}
\end{figure*}
%%%%
\begin{figure*}
 \includegraphics[width=\textwidth]{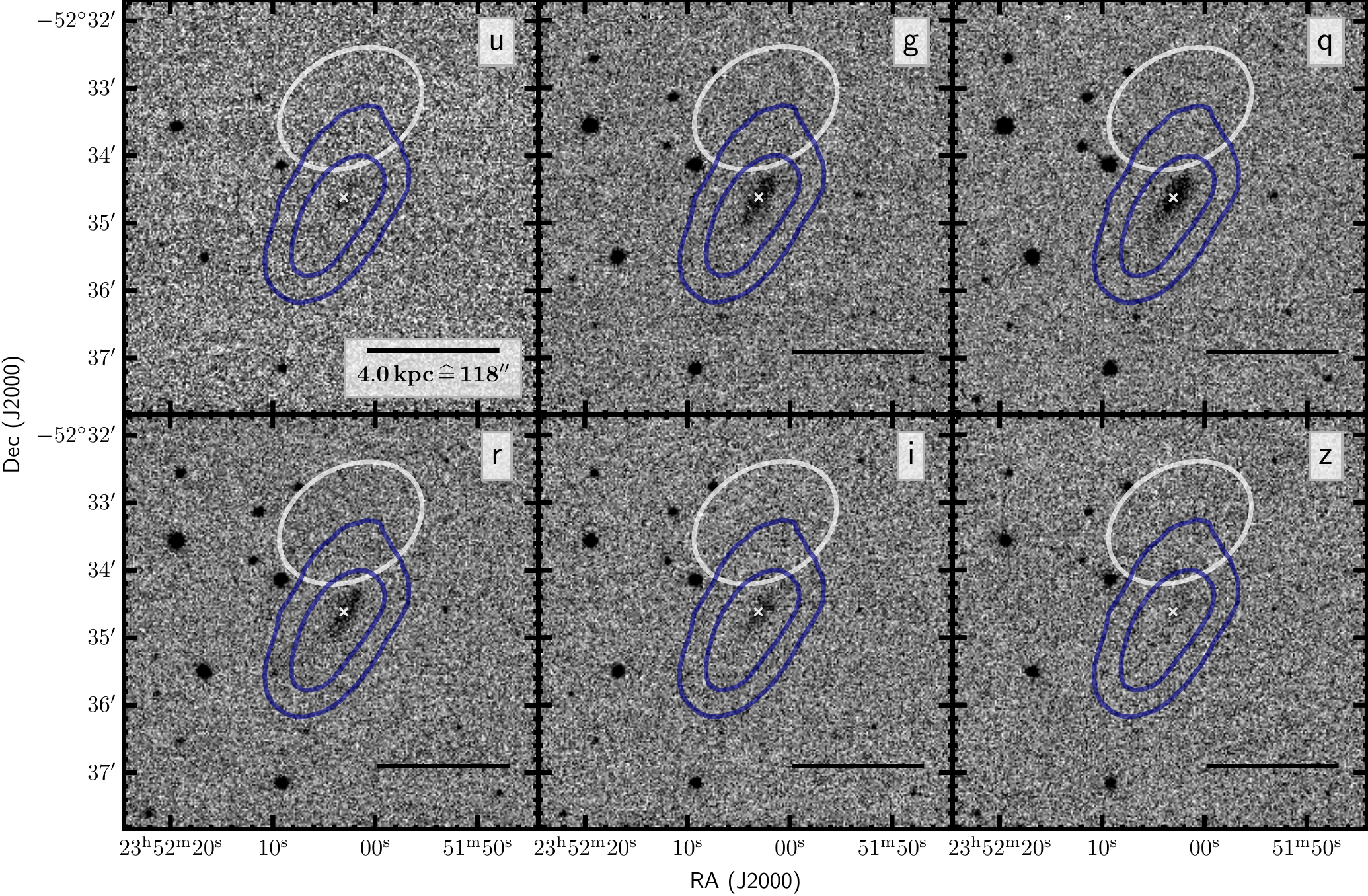}
 \caption{MeerLICHT observations of ESO~149-G003. Contours denote the $1,4\,$-$\,M_\odot\,\mathrm{pc}^{-2}$-levels of the observed total-intensity map shown in Figs.~\ref{fig:ESO149-G003_mom0_mom1_vf} and \ref{fig:ESO149-G003_mom0_mom1_vf_obs}. The white ellipse to the top highlights the position of kinematically anomalous \ion{H}{i}.
 }
 \label{fig:ESO149-G003_optical_meerlicht}
\end{figure*}
% Don't change these lines
\bsp	% typesetting comment
\label{lastpage}
\end{document}